%% file: main.tex
\documentclass[11pt]{article}

\usepackage[final]{acl}

\usepackage{times}
\usepackage{latexsym}
\usepackage{subcaption}
\usepackage[T1]{fontenc}
\usepackage{hyperref}
\usepackage[utf8]{inputenc}
\usepackage{booktabs}
\usepackage{multirow}
\usepackage{colortbl}
\usepackage{xcolor}
\usepackage{microtype}
\usepackage{multirow}

\usepackage{inconsolata}
\usepackage{amsmath}
\usepackage{graphicx}
\usepackage{booktabs}
\usepackage{pifont}
\usepackage[most]{tcolorbox}
\tcbuselibrary{breakable}
\newcommand{\cmark}{\ding{51}}
\newcommand{\xmark}{\ding{55}}
\definecolor{back_color}{gray}{0.95} 
\definecolor{easy_color}{HTML}{2e4057}
\definecolor{medium_color}{HTML}{083d77}
\definecolor{hard_color}{HTML}{DA4167}
\definecolor{traj_color}{HTML}{3092BF}
\definecolor{oursrow}{RGB}{233,243,253}   
\definecolor{oraclerow}{RGB}{255,243,230} 
\definecolor{headergray}{gray}{0.94}
\title{StressWeb: A Diagnostic Benchmark for Web Agent Robustness \\ under Realistic Interaction Variability}

\author{
\textbf{Haoyue Bai\textsuperscript{1}\thanks{Equal contribution}},
\textbf{Dong Wang\textsuperscript{1}\footnotemark[1]},
\textbf{Long Chen\textsuperscript{1}},
\textbf{Bingguang Hao\textsuperscript{1}}\\
\textbf{Pengyang Shao\textsuperscript{2}},
\textbf{Yonghui Yang\textsuperscript{2}},
\textbf{Yicheng He\textsuperscript{3}},
\textbf{Chenyi Zhuang\textsuperscript{1}
}\\
\\
\textsuperscript{1}Inclusion AI, Ant Group\\
\textsuperscript{2}National University of Singapore\\
\textsuperscript{3}University of Illinois Urbana-Champaign
}

\begin{document}
\maketitle

\input{section/abstract}
\input{section/introduction}
\input{section/relatedWork}

\input{section/methodology}
\input{section/experiment}
\input{section/conclusion}
\newpage
\bibliography{custom}
\input{section/appendix}

\end{document}

%% file: section/abstract.tex
\begin{abstract}
Large language model-based web agents have demonstrated strong performance on realistic web interaction tasks. 
However, existing evaluations are predominantly conducted under relatively stable and well-behaved interaction conditions, which may overestimate agent robustness. 
High task success in such idealized settings does not necessarily reflect performance under realistic web interaction. 
To address this limitation, we introduce a diagnostic stress-testing benchmark for web agents. 
We first construct realistic and controllable web environments that provide clean and stable interaction workflows as reference baselines. 
We then introduce structured and controlled perturbations that emulate interaction variability, including shifting layouts, altered interaction semantics, and execution disruptions.
By comparing agent behavior between clean and perturbed settings, our framework enables systematic diagnosis of robustness under what-if interaction scenarios.
Through extensive evaluation of state-of-the-art multimodal web agents, we show that stress-based evaluation exposes failure modes and substantial robustness gaps that remain hidden under clean benchmark conditions.
\end{abstract}

%% file: section/introduction.tex
\section{Introduction}

\begin{figure}[ht]
    \centering
    \includegraphics[width=0.48\textwidth]{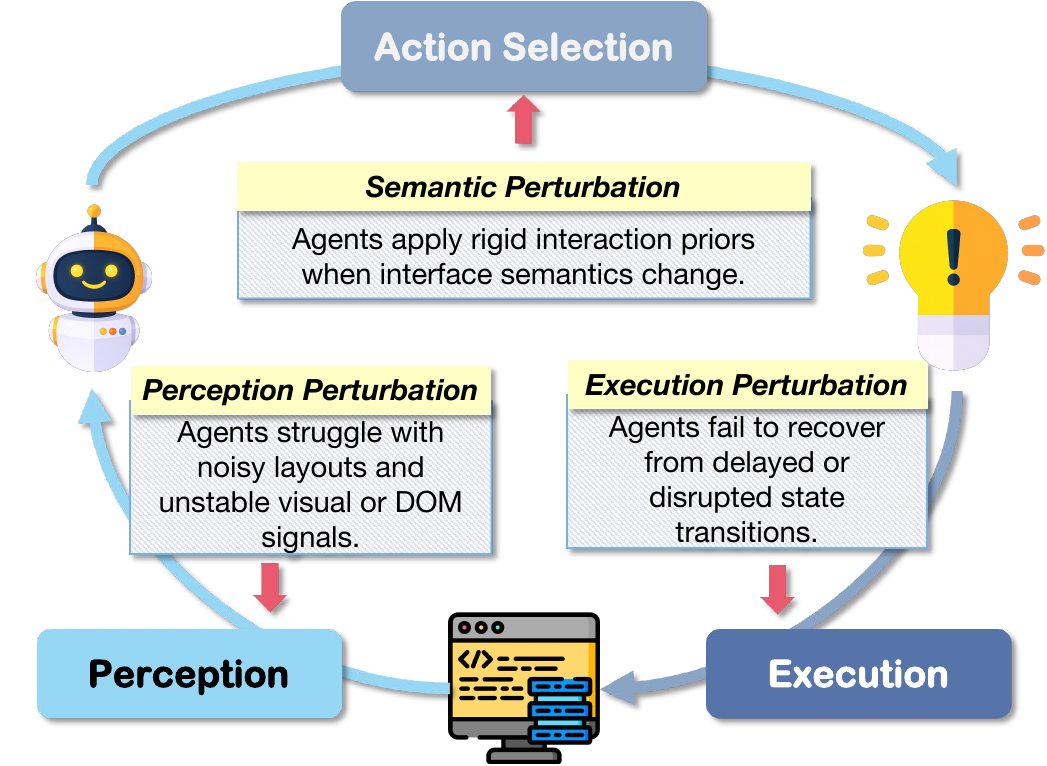}
    \caption{\small{Perturbations Across the Interaction Pipeline}}
    \label{fig:intro}
\end{figure}

Large language model (LLM)--based web agents have recently achieved strong performance on complex web interaction tasks such as form filling, online shopping, and information management~\cite{deng2023mind2web, ning2025survey,liu2025debate}. 
These agents integrate multimodal perception and sequential decision-making to interact directly with complex web interfaces.
\footnote{\url{https://github.com/inclusionAI/AWorld-RL/tree/main/StressWeb}}
\footnote{For correspondence: \texttt{baihaoyue621@gmail.com}.}

Alongside these advances, increasingly realistic benchmarks have been developed to evaluate agent performance in environments that resemble real-world websites.
Platforms such as WebArena~\cite{zhou2023webarena} and WebVoyager~\cite{he2024webvoyager} deploy full-stack websites to improve ecological validity, while works such as REAL~\cite{garg2025real} and MLA-Trust~\cite{yang2025mla} emphasize simulation and reproducibility.
However, despite increasing realism, evaluation is still largely conducted under relatively stable and well-behaved interaction conditions.

In real-world settings, web interfaces frequently exhibit asynchronous responses, delayed or failed updates, unexpected pop-up interruptions, shifting layouts, and evolving interaction semantics~\cite{violations1,violations2,violations3,danger,popup,dompert}.
Such variability challenges agent perception, action selection, and execution in ways that are rarely stress-tested in standard benchmarks.
Consequently, high task success in idealized benchmark environments does not necessarily reflect robustness under dynamic and irregular web interaction.

To address this gap, we propose a diagnostic stress-testing benchmark for web agents.
Rather than passively measuring performance in fixed environments, we introduce controlled perturbations that simulate realistic interaction variability.
This intervention-based design enables a structured analysis of agent robustness under irregular conditions.
We operationalize this framework by first constructing a collection of clean, realistic web environments that provide stable and comprehensive interaction workflows across diverse task types.
These environments serve as reliable reference baselines.
We then derive perturbed counterparts by injecting carefully designed stage-aligned interference into the interaction process (Figure~\ref{fig:intro}):
(1) perception perturbations that introduce visual and DOM irregularities,
(2) semantic perturbations that alter action semantics under explicit or implicit cues, and 
(3) execution perturbations that induce action failures or disruptive interruptions.
Each task is paired with a deterministic multi-checkpoint evaluation script that automatically verifies intermediate interaction milestones as well as final outcomes.

Through extensive evaluation of multimodal LLM web agents, we show that stress perturbations expose failure modes hidden under clean evaluation settings.
Our results reveal substantial gaps between stable benchmark performance and behavior under dynamic interaction variability, highlighting the importance of intervention-based evaluation.
In summary, this work makes three contributions:
\begin{itemize}
    \item We introduce a diagnostic stress-testing benchmark that evaluates web agent robustness under irregular interaction conditions.
    \item We design stage-aligned perturbation mechanisms that enable robustness analysis across perception, action selection, and execution.
    \item Our stress-based evaluation reveals hidden failure modes and robustness gaps concealed by idealized benchmark settings.
\end{itemize}

%% file: section/relatedWork.tex
\section{Related Work}
\label{sec:related_work}

\subsection{Web Agent Benchmarks: From Simplified to Realistic Environments}

Early research on web agents focused on sandboxed environments that enable agents to interact with simplified web interfaces.
World of Bits~\cite{shi2017world} introduced one of the first open-domain platforms for web-based agents, covering tasks such as form filling and navigation across diverse websites.
Subsequently, WebShop~\cite{yao2022webshop} proposed a scalable simulated e-commerce environment to evaluate grounded language agents on goal-oriented shopping tasks.
Recent benchmarks have significantly advanced the realism of web agent evaluation.
Mind2Web~\cite{deng2023mind2web} constructs a large-scale dataset of real-world web interaction trajectories, enabling supervised training and evaluation of generalist web agents.
WebArena~\cite{zhou2023webarena} deploys full-stack realistic websites in controlled environments, enabling reproducible evaluation while preserving real-world complexity.
Similarly, WebVoyager~\cite{he2024webvoyager} demonstrated an end-to-end web agent that operates directly on screenshots and natural language instructions, highlighting the potential of large multimodal models for realistic web interaction.
More recently, BrowserArena~\cite{anupam2025browserarena} evaluates agents on realistic web navigation tasks across diverse websites, further emphasizing end-to-end interaction capabilities of modern LLM agents.
In parallel, several recent benchmarks evaluate browsing and information-seeking capabilities of web agents.
BrowseComp~\cite{wei2025browsecomp} introduces a challenging benchmark focused on multi-step web browsing and information retrieval tasks.
MM-BrowseComp~\cite{li2025mmbrowsecomp} further extends this setting to multimodal browsing scenarios where agents must reason over visual and textual web content.
Similarly, DeepShop~\cite{lyu2025deepshop} and WebMall~\cite{peeters2025webmall} study shopping-oriented web agents that retrieve and compare product information across multiple online stores.
Despite these advances, most existing benchmarks remain largely static once instantiated.
While increased realism improves ecological validity, it also introduces uncontrolled variability.
Consequently, failures are difficult to attribute to perception, planning, or incidental environmental factors (e.g., asynchronous loading or interface changes).
This limits their diagnostic value and motivates the need for environments that are both realistic and controllable.

\subsection{Simulators and Environment Modeling}

To address the tension between realism and controllability, several recent works have explored simulated web environments.
REAL~\cite{garg2025real} evaluates agents on deterministic simulations of real websites, enabling execution errors to be isolated by eliminating stochastic environmental factors.
WebCanvas~\cite{pan2024webcanvas} introduces an benchmark that emphasizes controllable environment replay while maintaining realistic interaction workflows.
Beyond evaluation, simulators have also been adopted as tools for improving agent robustness through training.
WebAgent-R1~\cite{wei2025webagent} leverages end-to-end multi-turn reinforcement learning in simulated web environments to enhance agent recovery behaviors.
Similarly, GUI Exploration Lab~\cite{yan2025gui} demonstrates that reinforcement learning in simulated GUI environments can significantly improve exploration efficiency and robustness in complex screen navigation tasks.
These works suggest a growing consensus that simulated environments are valuable for controlled evaluation and training.
However, prior approaches mainly use them for data generation or performance improvement.
In contrast, our work uses controllable web environments to enable diagnostic evaluation, introducing structured perturbations that systematically probe agent reliability under realistic interaction variability.

\subsection{Reliability and Robustness of LLM Agents}

As large language model agents are increasingly deployed in interactive environments, understanding their reliability and robustness has become a critical research challenge~\cite{shao2026baldro,zhai2026maximizing}.
Recent studies have moved beyond aggregate task success rates to analyze agent failures in greater depth.
WAREX~\cite{kara2025warex} performs a large-scale reliability analysis of web agents across existing benchmarks.
By repeatedly evaluating agents under identical task settings, it reveals substantial performance variance and execution instability.
However, WAREX is not a benchmark or environment itself, but an analysis framework operating on fixed benchmarks, and thus relies on post-hoc outcome statistics without the ability to intervene in or manipulate environment conditions.
In parallel, MLA-Trust~\cite{yang2025mla} benchmarks the trustworthiness of multimodal LLM agents in GUI environments, focusing on perception, reasoning, and safety-related failures.
While these efforts provide valuable insights into agent reliability, they primarily analyze failures under predefined task distributions and static environment configurations.
In contrast, our work adopts an interventional perspective on reliability evaluation.
By actively injecting controlled perturbations into the web environment, we enable validation of agent failures.
This approach moves beyond correlational post-hoc analysis toward mechanistic understanding of how and why agents fail under specific environmental disruptions, complementing existing benchmarks and reliability analyses. A comparison with existing benchmarks
is provided in Appendix~\ref{sec:appendix_benchmark_comparison}.

%% file: section/methodology.tex
\section{Methodology}

\begin{figure*}[ht]
    \centering
    \includegraphics[width=0.98\textwidth]{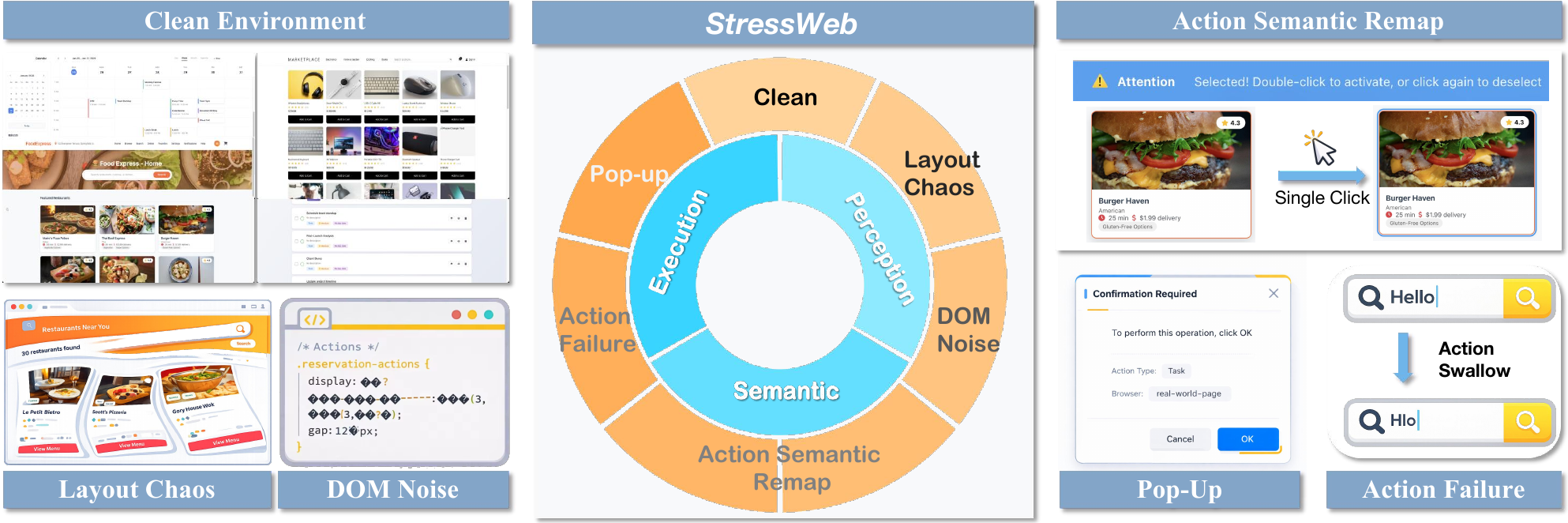}
    \caption{{Overview of the StressWeb benchmark. 
Clean reference environments provide stable interaction workflows as reference.
The benchmark introduces six perturbation types spanning perception (layout chaos, DOM noise), action selection (action semantic remap with or without explicit cues), and execution (pop-up interruptions, action failures) to systematically evaluate web agent robustness under controlled interaction variability.}}
    \label{fig:Overview}
\end{figure*}

\subsection{Benchmark Overview}
\label{sec:benchmark_overview}
Our benchmark evaluates web agent robustness under controlled environmental stress.
It consists of clean reference environments, structured perturbations, and deterministic multi-checkpoint evaluators (Figure~\ref{fig:Overview}).

To systematically organize perturbations, we view web agent interaction as a structured pipeline rather than a single atomic transition.
Each interaction cycle can be decomposed into three stages:
(1) perception, where the agent observes the current interface;
(2) action selection, where the agent selects an action based on perceived content and learned action semantics; and
(3) execution, where the environment realizes the issued action and updates its internal state.
Perceptual perturbations modify state exposure, semantic perturbations alter action selection rules, and execution perturbations affect transition reliability.
Each task is first evaluated in a clean setting, where the interaction pipeline is stable and deterministic across all three stages.
The same tasks are then re-evaluated under controlled perturbations with instructions fixed, enabling diagnostic comparison across perception, semantic, and execution dimensions.

\subsection{Reference Environments}
\label{sec:clean_env}

To enable controlled stress testing, we construct a set of clean reference web environments that serve as the baseline reference for all subsequent evaluations.
In these environments, all stages of the interaction pipeline, including perception, action selection, and execution, operate in a stable and fully deterministic manner.
This provides a reliable baseline against which the effects of structured perturbations can be clearly isolated and measured.

\paragraph{Construction Pipeline.}
Each environment is generated through an agent-driven website construction pipeline that converts natural language task specifications into fully functional web applications.
Unlike static template-based generation, the pipeline produces structurally coherent websites with realistic navigation flows, persistent internal state, and task-consistent data.
The controllable simulation framework also makes the interaction dynamics remain stable and reproducible~\cite{garg2025real}.
The generation process is automated yet controllable, enabling scalable environment expansion and the construction of training data that is distributionally aligned with evaluation settings.
To ensure reliability, all generated websites undergo manual verification and light refinement to confirm functional correctness and interface stability.
Implementation details are provided in Appendix~\ref{app:task_driven_construction}.

\paragraph{Benchmark Scope.}
We construct $10$ websites covering different applications comprising $149$ task queries (more details in Appendix~\ref{sec:appendix_websites}).
Each query is evaluated under $7$ environment settings, resulting in a total of $1{,}043$ task evaluations.
The benchmark covers representative end-to-end interaction patterns, including multi-step navigation, form completion, search and filtering, state modification, and batch operations.
These tasks require consistent perception, coherent action selection, and reliable execution across multiple steps, reflecting common real-world interaction scenarios.

\subsubsection{Perception Perturbations}
\label{sec:perc_perturb}

Perception perturbations target the perception stage of the interaction pipeline.
At this stage, the agent observes the rendered interface and identifies task-relevant elements from visual and structural signals.
In real-world web systems, the observable interface often contains visual noise or structural irregularities due to legacy code, dynamic rendering, and anti-crawling mechanisms.
Such conditions can disrupt agents that rely on clean or template-like interface representations.
To evaluate robustness at this stage, we construct perception-perturbed environments that introduce controlled visual and structural variability.
These environments require agents to remain robust to noise and to correctly identify task-relevant elements despite misleading or redundant interface signals.

Concretely, we construct two distinct perception-perturbed environments: a visual-perturbed environment  and a DOM-perturbed environment.
In the visual-perturbed setting (Chaos), we modify the rendered presentation of interface elements through changes such as altered text size, element rotation, and positional shifts.
In the DOM-perturbed setting (Noise), we introduce structural redundancy and lightweight obfuscation patterns that resemble common real-world practices.
These include text fragmentation, hidden or decoy elements, attribute perturbation, and mild content encoding~\cite{sarabamoun2025special}.
Perception perturbations modify how the underlying state is exposed to the agent, testing robustness under noisy or structurally misleading interface signals. The intensity of both visual and structural perturbations is parameterizable.
Geometric distortion magnitude, noise density, and structural redundancy levels can be systematically adjusted to simulate different degrees of perceptual stress, enabling graded evaluation rather than a single fixed perturbation setting.

\subsubsection{Semantic Perturbations}
\label{sec:sem_perturb}

Semantic perturbations target the action selection stage of the interaction pipeline.
At this stage, the agent selects actions based on learned assumptions about interface semantics, such as single-click for navigation or confirmation.
In real-world web systems, however, interaction rules are often implicit or deviate from common conventions.
When such deviations occur, agents that rigidly apply default action priors may systematically fail~\cite{sepert1}.
To evaluate robustness at this stage, we construct semantic-perturbed environments that modify the mapping between issued actions and their effective outcomes while keeping the observable interface unchanged.
These perturbations require agents to infer altered interaction rules from contextual cues and revise default action assumptions accordingly.

Concretely, we instantiate this design using two semantic-perturbed environments.
In both environments, selected single-click operations are remapped to require double-click execution. In the explicit environment (RemapE), the interface provides textual instructions indicating the modified interaction rule, requiring the agent to incorporate new semantic information into its decision process. In the implicit environment (Remap), no such instruction is provided; instead, a single click only produces a selection state, and the intended transition is triggered by a double click. The implicit environment therefore requires the agent to infer the altered rule through interaction feedback rather than relying on explicit guidance.

\subsubsection{Execution Perturbations}
\label{sec:exec_perturb}

Execution perturbations target the execution stage, where issued actions are realized and translated into state updates.
In real-world web systems, action execution may be unreliable due to asynchronous processing, transient failures, or unexpected interface interruptions.
As a result, agents cannot assume that every issued action immediately produces its intended effect.
To evaluate robustness at this stage, we construct execution-perturbed environments in which action realization becomes unreliable.
These perturbations require agents to verify execution outcomes and adapt their behavior when intended transitions do not occur.

Concretely, we design two execution-perturbed environments.
In the first environment (Failure), actions may randomly fail to execute, including both click operations and text input submissions.
In the second environment (Pop-Up), disruptive pop-up windows are introduced with varying prompts and interaction logic, requiring the agent to handle interruptions before continuing the intended task flow.
Both mechanisms are parameterizable, allowing the failure rate and interruption frequency to be adjusted to simulate different levels of execution stress.
This design enables systematic control over execution-stage variability while preserving the underlying task semantics.

\subsection{Deterministic Evaluation}
\label{sec:evaluation}

To ensure reliable and diagnostic assessment, we adopt a deterministic, script-based evaluation protocol that verifies agent behavior at multiple checkpoints throughout task execution.
Rather than relying solely on final task completion, this design enables fine-grained attribution of failures to specific interaction stages.

For each task, we define a set of predefined checkpoints corresponding to critical milestones, such as successful navigation to a target page, correct completion of required form fields, or proper triggering of an intended action.
After each agent action, the intermediate DOM state is recorded and automatically evaluated by task-specific checking scripts, which compare the state against expected conditions at that checkpoint.
By grounding evaluation in observable states rather than agent-reported success signals, the protocol ensures consistent, reproducible, and stage-aware assessment, particularly under perturbation settings where agents may partially succeed before deviating.

%% file: section/experiment.tex
\section{Experiments}
\label{sec:experiments}

For completeness, we provide additional experimental results, detailed analyses, and failed pattern analysis in Appendix~\ref{sec:appendix_detailed_results}, \ref{sec:appendix_self_assessment}, \ref{sec:appendix_model_ranking}, \ref{sec:discussion}, and \ref{sec:Repetitive_Action_Failure}.

\begin{table*}[t]
\centering

\renewcommand{\arraystretch}{1.5}
\setlength{\tabcolsep}{4.5pt}

\caption{{Overall Performance Across Different Environments. We report checkpoint pass rate (ckpt\%) and average interaction steps (steps).}}
\label{tab:model_performance}

\resizebox{\textwidth}{!}{
\begin{tabular}{lcccccccccccccccc}
\toprule

\multirow{2}{*}{\textbf{Model}}
& \multicolumn{2}{c}{\textbf{Clean}}
& \multicolumn{2}{c}{\textbf{Chaos}}
& \multicolumn{2}{c}{\textbf{Noise}}
& \multicolumn{2}{c}{\textbf{Failure}}
& \multicolumn{2}{c}{\textbf{Pop-Up}}
& \multicolumn{2}{c}{\textbf{RemapE}}
& \multicolumn{2}{c}{\textbf{Remap}}
& \multicolumn{2}{c}{\textbf{Avg}} \\

\cmidrule(lr){2-3}
\cmidrule(lr){4-5}
\cmidrule(lr){6-7}
\cmidrule(lr){8-9}
\cmidrule(lr){10-11}
\cmidrule(lr){12-13}
\cmidrule(lr){14-15}
\cmidrule(lr){16-17}

& ckpt\% $\uparrow$ & steps $\downarrow$
& ckpt\% $\uparrow$ & steps $\downarrow$
& ckpt\% $\uparrow$ & steps $\downarrow$
& ckpt\% $\uparrow$ & steps $\downarrow$
& ckpt\% $\uparrow$ & steps $\downarrow$
& ckpt\% $\uparrow$ & steps $\downarrow$
& ckpt\% $\uparrow$ & steps $\downarrow$
& ckpt\% $\uparrow$ & steps $\downarrow$ \\

\midrule

GLM-4.6V
& 46.7 & 61.0
& 38.3 & 59.2
& 42.1 & 61.6
& 40.1 & 59.3
& 38.0 & 66.6
& 20.8 & 89.1
& 20.5 & 88.4
& 35.2 & 69.3 \\

Qwen3-VL-235B
& 52.7 & 60.0
& 44.9 & 61.2
& 49.2 & 58.8
& 43.4 & 62.3
& 40.9 & 68.5
& 26.5 & 84.8
& 24.8 & 92.0
& 40.3 & 69.7 \\

Claude Haiku 4.5
& 43.0 & 79.8
& 38.5 & 77.3
& 40.9 & 83.1
& 39.7 & 83.1
& 38.5 & 85.8
& 26.0 & 95.0
& 24.1 & 97.7
& 34.6 & 87.0 \\

Claude Opus 4.5
& 55.0 & 66.2
& 41.6 & 71.1
& 51.5 & 67.0
& 47.7 & 71.5
& 42.3 & 80.5
& 39.0 & 85.7
& 28.9 & 94.0
& 43.7 & 76.6 \\

Claude Sonnet 4.5
& 57.5 & 64.4
& 45.7 & 74.6
& 52.4 & 67.9
& 47.5 & 72.7
& 45.2 & 74.6
& 31.6 & 89.8
& 25.4 & 90.1
& 43.6 & 76.3 \\

Gemini-2.5-Pro
& 59.3 & 49.5
& 53.7 & 51.4
& 55.2 & 54.2
& 51.3 & 56.2
& 49.0 & 67.1
& 39.4 & 79.1
& 28.9 & 84.9
& 48.1 & 63.2 \\

GPT-5.2
& 55.5 & 66.4
& 49.1 & 63.6
& 49.6 & 66.4
& 45.5 & 72.2
& 39.8 & 78.1
& 29.0 & 93.1
& 23.4 & 95.7
& 41.7 & 76.5 \\

o4-mini
& 61.4 & 48.6
& 54.8 & 51.5
& 57.9 & 49.1
& 54.0 & 54.8
& 50.3 & 65.9
& 38.3 & 79.4
& 33.7 & 82.4
& 50.1 & 61.7 \\

\midrule

\rowcolor{oraclerow}
Human (Oracle)
& 100 & 7.0
& 100 & 7.0
& 100 & 7.0
& 100 & 10.3
& 100 & 9.7
& 100 & 7.8
& 100 & 7.7
& 100 & 8.1 \\

\bottomrule
\end{tabular}
}

\end{table*}



\subsection{Experimental Setup}
\label{sec:exp_setup}

\paragraph{Models.}
We evaluate a diverse set of state-of-the-art multimodal and reasoning-capable
LLM systems accessed via public APIs, including the
Gemini family (Gemini-2.5-Pro),
OpenAI GPT series (GPT-5.2, o4-mini),
Anthropic Claude series (Claude-Sonnet-4.5, Claude-Opus-4.5, Claude-Haiku-4.5),
GLM-V models (GLM-4.6V),
and Qwen-VL models (Qwen3-VL-235B-A22B-Thinking).
In addition, we include a human baseline to provide an upper-bound reference (conducted by computer science graduate students).

\paragraph{Run Protocol.}
All models interact with environments through a unified Playwright-based controller.
At each step, the agent receives the current page screenshot and screenshots from the two most recent, the page DOM representation in HTML format, and the full history of previous actions and execution outcomes.
Each run is limited to 100 interaction steps.
A run terminates when the agent outputs DONE or FAIL, or when the step budget is exhausted.
Task success is determined by task-specific evaluators that verify final page states against predefined criteria.
All models are evaluated under seven conditions (one clean setting and six perturbed variants) using the same task set and interaction protocol.
In all reported experiments, perturbations are instantiated at a moderate stress level, with execution failure probability set to 0.35 and visual/structural distortions calibrated to intermediate severity.
While this work focuses on a fixed stress configuration for controlled comparison, the benchmark supports parameterizable perturbation intensity.


\begin{figure*}[ht]
    \centering
    \includegraphics[width=0.98\textwidth]{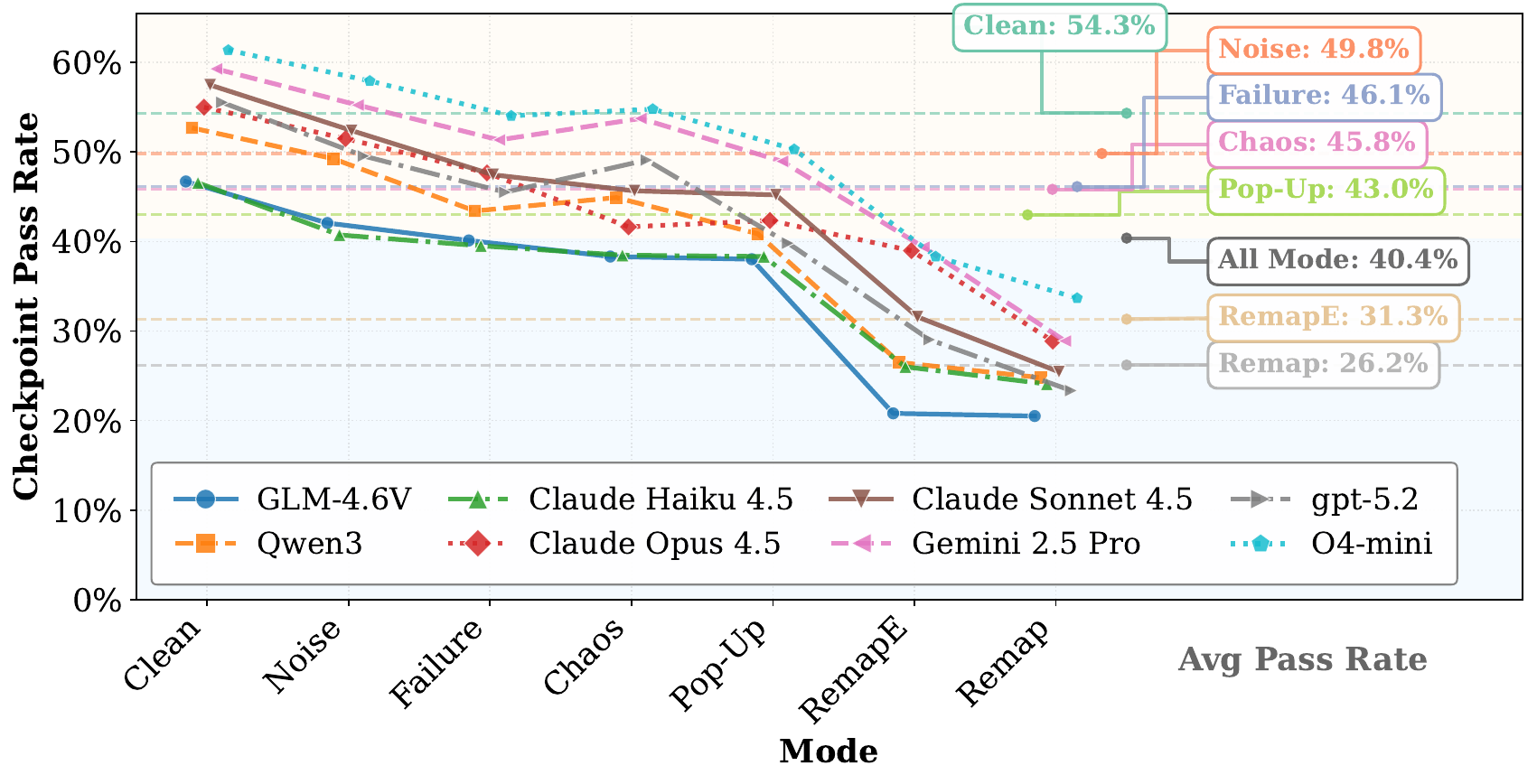}
\caption{{
Checkpoint pass rates across different perturbation modes.
Lines represent individual models and the dashed horizontal lines indicate the average performance for each mode across all models.
Perception perturbations (Noise and Chaos) introduce relatively mild degradation, while execution perturbations (Failure and Pop-Up) lead to moderate drops.
Semantic perturbations (RemapE and Remap), which alter interaction rules, cause the most severe performance decline.
}}
    \label{fig:mode_pass_rate}
\end{figure*}

\subsection{Overall Performance under Environmental Stress}

Table~\ref{tab:model_performance} summarizes the performance of representative multimodal models acting as web agents under clean and perturbed environments.
Performance is measured by checkpoint success rate (ckpt\%) and average interaction steps.
Under the clean setting, agents achieve success rates between 46.7\% and 61.4\%.
However, across perturbation settings, performance drops to 34.6\%--50.1\%, revealing a clear gap between capability and robustness.
While agents perform well under stable conditions, they fail to maintain performance when interaction assumptions are perturbed, suggesting reliance on fragile environment-specific priors rather than adaptive strategies.
Although o4-mini and Gemini-2.5-Pro achieve relatively higher average success rates (50.1\% and 48.1\%), all models exhibit substantial degradation, indicating that robustness limitations are systemic.

In contrast, the human oracle achieves near-perfect success across all environments with minimal increase in interaction steps.
This suggests that the perturbations do not fundamentally increase task difficulty, but instead expose limitations in how agents perceive, reason, and adapt to changing environments.
Therefore, the observed degradation reflects insufficient robustness and adaptability rather than intrinsic task complexity.

\subsection{Perturbation Sensitivity Analysis}
\label{sec:mode_analysis}

To understand the impact of environmental perturbations, 
Figure~\ref{fig:mode_pass_rate} reports checkpoint pass rates across models and perturbation modes.
Environmental perturbations consistently degrade performance across all systems, indicating high sensitivity to deviations from expected interaction patterns.
Compared with the clean environment, success rates decrease under every perturbation setting, and this trend is consistent across all models, suggesting that even moderate variability can undermine agent reliability.
Different perturbation categories exhibit distinct levels of difficulty.
Perception perturbations introduce relatively limited degradation, with average success decreasing from 54.3\% to 49.8\% under Noise and 45.8\% under Chaos.
Execution perturbations lead to moderately larger drops, to 46.1\% under Failure and 43.0\% under Pop-Up, indicating limited robustness to execution uncertainty and error recovery.

The most severe degradation occurs under semantic perturbations, where interaction rules deviate from common conventions.
Performance drops sharply to 31.3\% under RemapE and 26.2\% under Remap.
This suggests that agents rely on learned interaction priors rather than grounding actions in environment feedback.
When these priors are invalidated, agents fail to adapt, even after repeated interactions.
Notably, implicit remapping (Remap) causes more severe degradation than explicit remapping (RemapE), indicating difficulty in inferring updated interaction rules from context and limited capability for online adaptation.

Overall, these results reveal a hierarchy of failure modes: agents are relatively robust to perceptual noise, moderately affected by execution uncertainty, and highly vulnerable to semantic shifts that violate interaction assumptions.

\begin{figure*}[ht]
    \centering
    \begin{subfigure}{0.49\textwidth}
        \centering
        \includegraphics[width=\linewidth]{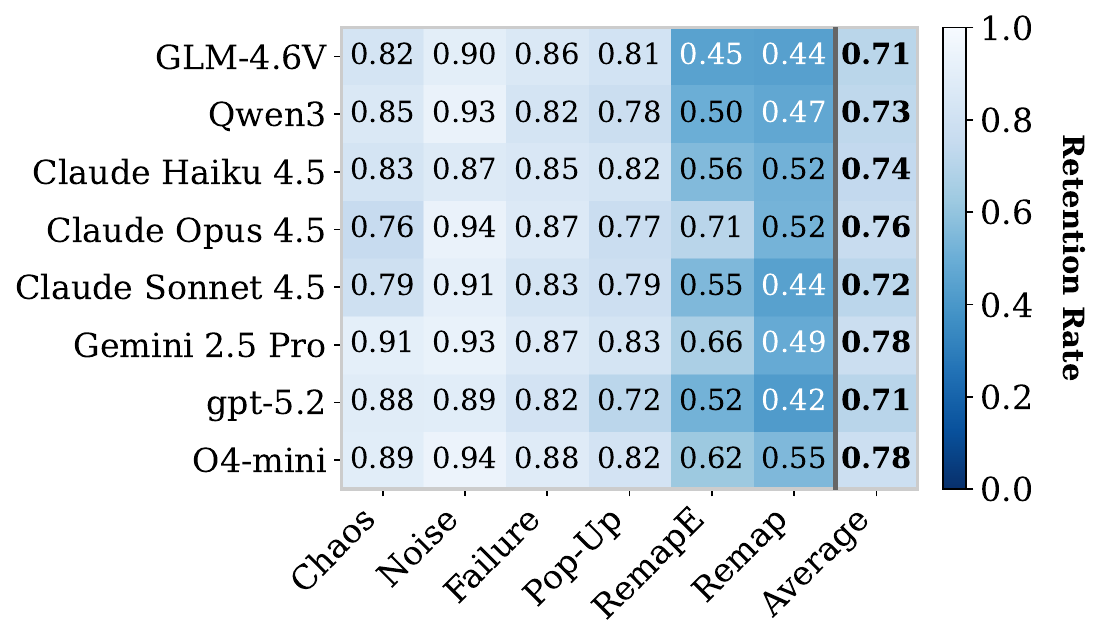}
        \caption{Performance Robustness}
        \label{fig:robustness_heatmap}
    \end{subfigure}
    \hfill
    \begin{subfigure}{0.49\textwidth}
        \centering
        \includegraphics[width=\linewidth]{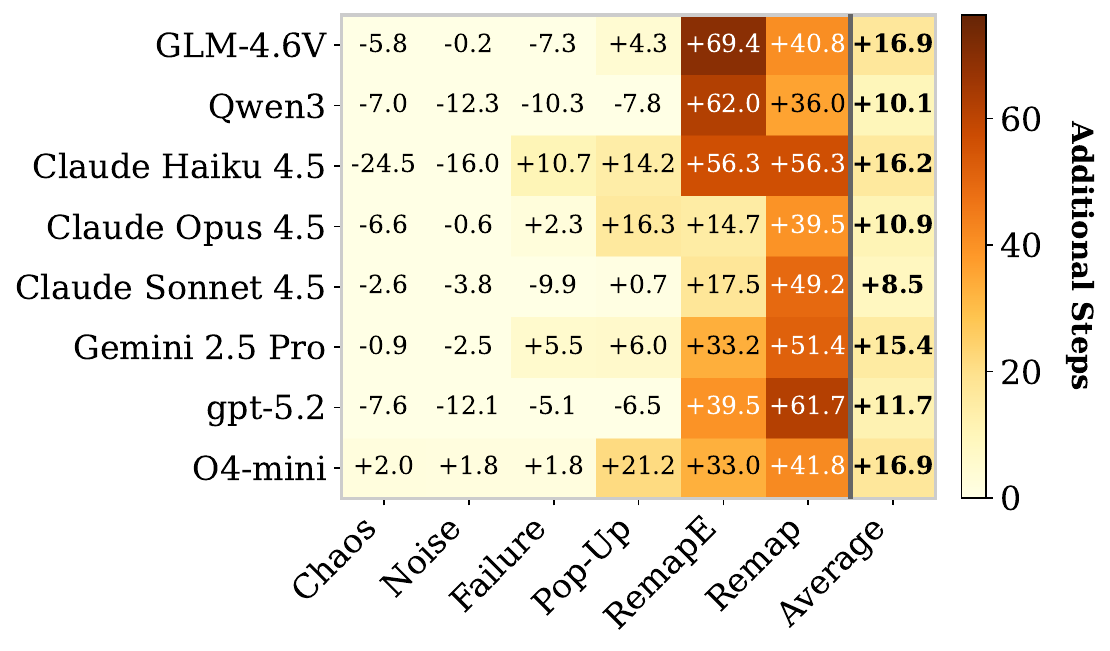}
        \caption{Step Difference}
        \label{fig:step_difference_heatmap}
    \end{subfigure}
    \caption{{Performance and cost robustness under environmental perturbations.
(a) Performance retention relative to the clean environment.
(b) Additional interaction steps compared to the clean setting.
Darker colors indicate stronger degradation or higher cost.}}
    \label{fig:heatmap}
\end{figure*}

\subsection{Performance and Interaction Cost Robustness}
\label{sec:robustness_heatmap}

Figure~\ref{fig:robustness_heatmap} reports performance retention relative to the clean environment, while Figure~\ref{fig:step_difference_heatmap} shows the additional interaction steps required under each perturbation.
Across all models, semantic perturbations again produce the most severe degradation.
When interaction rules are remapped (RemapE and Remap), performance retention drops dramatically.
For example, GLM retains only 45\% and 44\% of its clean performance under RemapE and Remap, respectively, while GPT-5.2 drops to 52\% and 42\%.
These results suggest that current web agents rely heavily on learned interaction conventions and struggle when action semantics change.
In contrast, perception and execution perturbations have noticeably milder effects.
Under Chaos, Noise, Failure, and Pop-Up conditions, most models retain a large portion of their clean performance, generally remaining close to or above 80\%.
Gemini-2.5-Pro and o4-mini show slightly stronger robustness in these settings, maintaining retention rates close to or above 0.85 across most non-semantic modes.
Claude Opus 4.5 also demonstrates comparatively strong robustness under semantic remapping.

The step analysis provides a complementary perspective on interaction cost.
While most perturbations increase the number of interaction steps, the largest increases occur under semantic perturbations.
For instance, GPT-5.2 requires over 60 additional steps under Remap, and several other models exhibit increases exceeding 50 steps.
Interestingly, performance degradation does not always correspond to longer interaction trajectories.
We observe two distinct failure patterns.
In some cases, agents terminate prematurely with fewer steps than in the clean environment, indicating early-stage failures caused by incorrect action assumptions.
In other cases, agents exhibit substantially longer trajectories, repeatedly attempting invalid actions without adapting their strategies.
For example, GLM and GPT-5.2 terminate earlier than in the clean environment, suggesting that agents commit to incorrect action hypotheses early and fail to recover.

Importantly, semantic perturbations not only reduce success rates but also substantially increase interaction cost, indicating that agents fail both to identify correct actions and to revise incorrect ones.
This dual failure highlights a key limitation: current agents lack mechanisms for online adaptation, leading to both incorrect decisions and inefficient exploration under changing interaction semantics.

\begin{figure}[ht]
    \centering
    \includegraphics[width=0.48\textwidth]{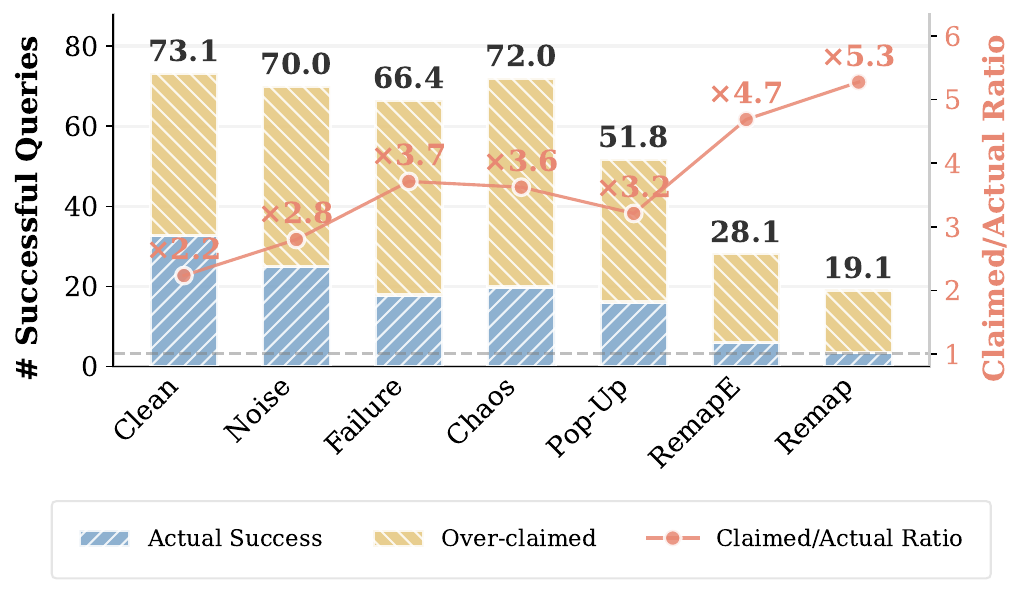}
    \caption{{Claimed vs. actual successful queries under different modes. 
Environmental perturbations increase the discrepancy between perceived and actual task success, indicating reduced self-assessment reliability.}}
    \label{fig:success_analysis_average}
\end{figure}

\subsection{Self-Assessment Reliability under Perturbations}
\label{sec:self_assessment}

Beyond task success, it is critical to examine whether agents can reliably assess their own outcomes.
In many web-agent frameworks, the model terminates an interaction by declaring task completion.
If this self-assessment is unreliable, agents may prematurely stop even when the task has not been correctly executed.
Figure~\ref{fig:success_analysis_average} compares the number of actually successful queries with those claimed as successful by the agent.
Even in the clean environment, agents exhibit noticeable miscalibration between claimed and actual task success.
Perturbations substantially amplify this miscalibration, leading to systematic overestimation of task success.
This suggests that agents lack reliable internal signals to detect failure under uncertain conditions.
The effect becomes particularly pronounced under semantic perturbations.
While the number of truly successful queries drops sharply under RemapE and Remap conditions, agents continue to report comparatively high success rates.
As a result, the claimed-to-actual success ratio increases significantly, reaching up to $5.3\times$, indicating severe overconfidence.
This behavior is consistent with the previously observed failure to adapt action strategies, indicating that agents not only make incorrect decisions but also fail to recognize these errors.
These results indicate that environmental shifts impair not only task execution but also agents’ ability to accurately evaluate their own performance.
Such miscalibration poses a critical challenge for real-world deployment, as agents may prematurely terminate tasks while confidently reporting success, leading to silent failures.

%% file: section/conclusion.tex
\section{Conclusion}

We introduced StressWeb, a diagnostic benchmark for evaluating the robustness of web agents under realistic interaction variability. 
By constructing controllable web environments with structured perturbations across perception, action selection, and execution stages, StressWeb enables systematic analysis of agent behavior under environmental stress.
Experiments on state-of-the-art multimodal agents reveal substantial robustness gaps that remain hidden under clean benchmark conditions. 
In particular, semantic perturbations that alter interaction rules cause the most severe performance degradation, while environmental stress also significantly impairs agents' ability to correctly assess their own task success. 

\section*{Limitations}

Despite enabling controlled robustness evaluation of web agents,
StressWeb has several limitations.
First, although the benchmark includes representative websites
from multiple task domains, the overall coverage of website
types remains limited compared with the diversity of real-world
web ecosystems. Expanding the benchmark to additional websites
and task categories would further improve its representativeness.
Second, the current perturbation mechanisms are primarily applied
individually to isolate their effects. In practice, multiple forms
of environmental variability may occur simultaneously.
Studying compositional perturbations and their combined effects
remains an important direction for future work.
Third, the perturbations used in StressWeb are controlled
abstractions designed to approximate realistic web variability.
While this design enables systematic analysis, real-world web
environments may exhibit more complex and dynamic behaviors
that are not fully captured in the current benchmark.

%% file: section/appendix.tex
\appendix
\begin{table*}[t]
\centering
\small
\caption{Comparison of representative web agent benchmarks and evaluation frameworks. A capability is marked as supported only if it is explicitly designed and systematically supported. \textbf{Actionable} indicates whether the benchmark requires agents to perform structured UI operations such as clicking, typing, and multi-step task workflows, rather than only browsing or reading web pages. \textbf{Realistic} denotes whether the environment preserves the fidelity of real-world web interfaces. \textbf{Deterministic} indicates whether task execution is reproducible across runs. \textbf{Diagnostic} denotes built-in support for attributing agent failures to specific components. \textbf{Interventional} indicates whether the benchmark supports controlled and reproducible environment perturbations for robustness evaluation.}
\begin{tabular}{lccccc}
\toprule
Benchmark &
Actionable &
Realistic &
Deterministic &
Diagnostic &
Interventional \\
\midrule
World of Bits & \cmark & \xmark & \cmark & \xmark & \xmark \\
WebShop & \cmark & \xmark & \cmark & \xmark & \xmark \\
Mind2Web & \cmark & \cmark & \xmark & \xmark & \xmark \\
WebArena & \cmark & \cmark & \cmark & \xmark & \xmark \\
WebVoyager & \cmark & \cmark & \xmark & \xmark & \xmark \\
BrowserArena & \cmark & \cmark & \xmark & \cmark & \xmark \\
\midrule
BrowseComp & \xmark & \cmark & \xmark & \xmark & \xmark \\
MM-BrowseComp & \xmark & \cmark & \xmark & \xmark & \xmark \\
DeepShop & \xmark & \cmark & \xmark & \xmark & \xmark \\
WebMall & \cmark & \cmark & \cmark & \xmark & \xmark \\
\midrule
REAL & \cmark & \cmark & \cmark & \xmark & \xmark \\
WebCanvas & \cmark & \cmark & \cmark & \xmark & \xmark \\
\midrule
\textbf{Ours} & \cmark & \cmark & \cmark & \cmark & \cmark \\
\bottomrule
\end{tabular}
\label{tab:benchmark_comparison}
\end{table*}

\section{Comparison with Existing Web Agent Benchmarks}
\label{sec:appendix_benchmark_comparison}

To provide a broader perspective on the design space of web agent benchmarks, 
Table~\ref{tab:benchmark_comparison} summarizes key properties of representative benchmarks and evaluation frameworks.

The comparison highlights several important dimensions that affect the evaluation of web agents.
First, actionable benchmarks require agents to perform structured UI operations 
such as clicking, typing, and multi-step task workflows, rather than only browsing or reading web pages.
Second, realistic environments preserve the fidelity of real-world web interfaces.
Third, deterministic environments ensure reproducible execution across runs.
Fourth, diagnostic benchmarks provide mechanisms for attributing agent failures to specific components.
Finally, intervention environments support controlled perturbations of environment dynamics, enabling causal analysis of agent robustness.

As shown in Table~\ref{tab:benchmark_comparison}, existing benchmarks typically emphasize only a subset of these properties.
Interactive benchmarks such as WebArena and WebVoyager prioritize realism but lack deterministic execution and diagnostic capabilities.
Browsing-oriented benchmarks such as BrowseComp and MM-BrowseComp focus primarily on information retrieval and do not require structured UI interaction.
Simulation-based environments such as REAL and WebCanvas improve reproducibility but provide limited mechanisms for systematic intervention.

In contrast, our benchmark is designed to jointly support actionable interaction, realistic environments, deterministic execution, diagnostic analysis, and controlled environmental perturbations.
This combination enables systematic evaluation of web agent robustness under controlled conditions.

\section{Task-Driven Website Construction}
\label{app:task_driven_construction}

To support controlled and reproducible evaluation of web agents, we construct benchmark environments using a task-driven website generation methodology.
Rather than starting from pre-existing websites and retrofitting evaluation tasks, our approach inverts this process: tasks are specified first, and websites are then constructed to faithfully support those tasks.
This design choice ensures that navigation structure, interaction semantics, and internal state transitions are explicitly aligned with the evaluation objectives.

\subsection{Task Decomposition and Specification}
\label{app:task_decomposition}

We begin by decomposing each benchmark scenario into a set of concrete, executable tasks.
Tasks are derived from high-level product requirements and are designed to reflect realistic user goals, such as completing a purchase, managing a document, scheduling events, or updating persistent records.
Importantly, all tasks are specified with explicit inputs and observable outcomes, avoiding underspecified or purely informational queries.

Each task is further decomposed into a sequence of interaction steps spanning multiple pages.
This decomposition makes explicit the required navigation paths, intermediate state changes, and final success conditions.
By grounding task definitions in step-level interaction structure, we ensure that task difficulty arises from environment behavior rather than ambiguity in task formulation.

\subsection{Task-Driven Website Construction}
\label{app:task_driven_websites}

Given a set of decomposed tasks, we construct web environments that are explicitly designed to support their execution.
Each website is generated to include the pages, routes, data entities, and interaction affordances required by the task set.
This task-driven construction ensures that page transitions, form submissions, and state updates follow coherent and deterministic logic, reducing spurious failures caused by missing links, inconsistent routing, or incomplete state propagation.

Because websites are generated \emph{after} task specification, the resulting environments naturally encode valid navigation paths and interaction flows for all tasks.
This contrasts with approaches that define tasks post hoc on existing websites, where agents may encounter undefined behavior due to incomplete coverage of the website’s interaction space.
In our setting, every task has a well-defined execution path by construction.

\begin{figure*}[ht]
    \centering
    \includegraphics[width=0.98\textwidth]{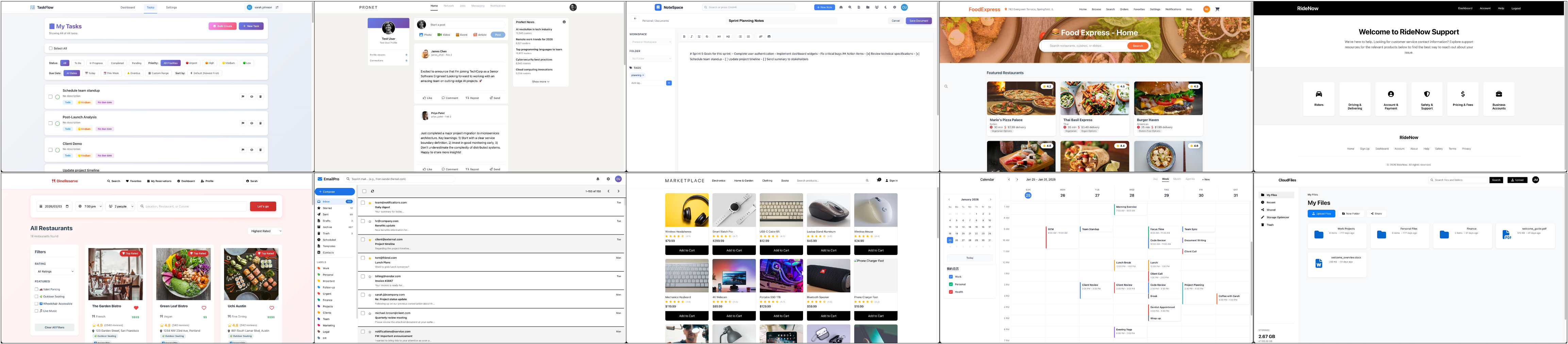}
    \caption{
Example screenshots of the generated web environments used in StressWeb.
The benchmark includes ten realistic website categories covering diverse domains such as e-commerce, scheduling, email management, transportation, and cloud file systems.
Each environment supports multi-step interactions, persistent state updates, and common web operations including search, filtering, navigation, and form submission.
}
    \label{fig:Wensites}
\end{figure*}

\subsection{Structured Task Trajectories}
\label{app:structured_trajectories}

To further bridge the gap between task specification and environment implementation, we represent each task as a structured trajectory.
A trajectory specifies the expected sequence of pages, user actions, and environment state transitions required for successful task completion.
These trajectories act as an intermediate representation connecting high-level task intent with low-level website behavior.

Structured trajectories serve two purposes.
First, they guide website construction by making explicit which pages, components, and backend operations must be implemented.
Second, they provide a reference specification for evaluation, enabling deterministic verification of intermediate and final outcomes.
This dual role ensures consistency between task design, website implementation, and evaluation criteria.

\subsection{From Trajectories to Executable Websites}
\label{app:website_generation}

Based on the structured task trajectories, we generate fully functional websites that implement the required frontend interfaces, backend logic, and persistent storage.
The generated websites support realistic interaction workflows, including multi-page navigation, form-based input, and stateful operations.
While the generation process is automated, it is constrained by the task specifications and trajectories, which act as a blueprint for correct behavior.

This process results in websites that are not merely visually plausible, but behaviorally aligned with the benchmark tasks.
As a consequence, agents evaluated in these environments are tested on robustness to environmental perturbations rather than on accidental inconsistencies or missing functionality.

\subsection{Human Verification and Refinement}
\label{app:human_verification}

Although the construction process is largely automated, we apply a lightweight human verification and refinement step to ensure suitability for benchmark evaluation.
This step focuses on verifying functional correctness of task workflows, clarity of interaction semantics, and consistency between task specifications and website behavior.
When discrepancies are identified—such as ambiguous UI cues, incomplete state updates, or minor inconsistencies in navigation—we apply minimal corrective edits.

Crucially, this refinement process is constrained to alignment and correctness fixes and does not introduce new functionality beyond the task-driven design.
As a result, the final environments remain faithful to the automated generation process while meeting the determinism and clarity requirements of reliable agent evaluation.

\begin{figure*}[t]
\centering
\includegraphics[width=0.22\textwidth]{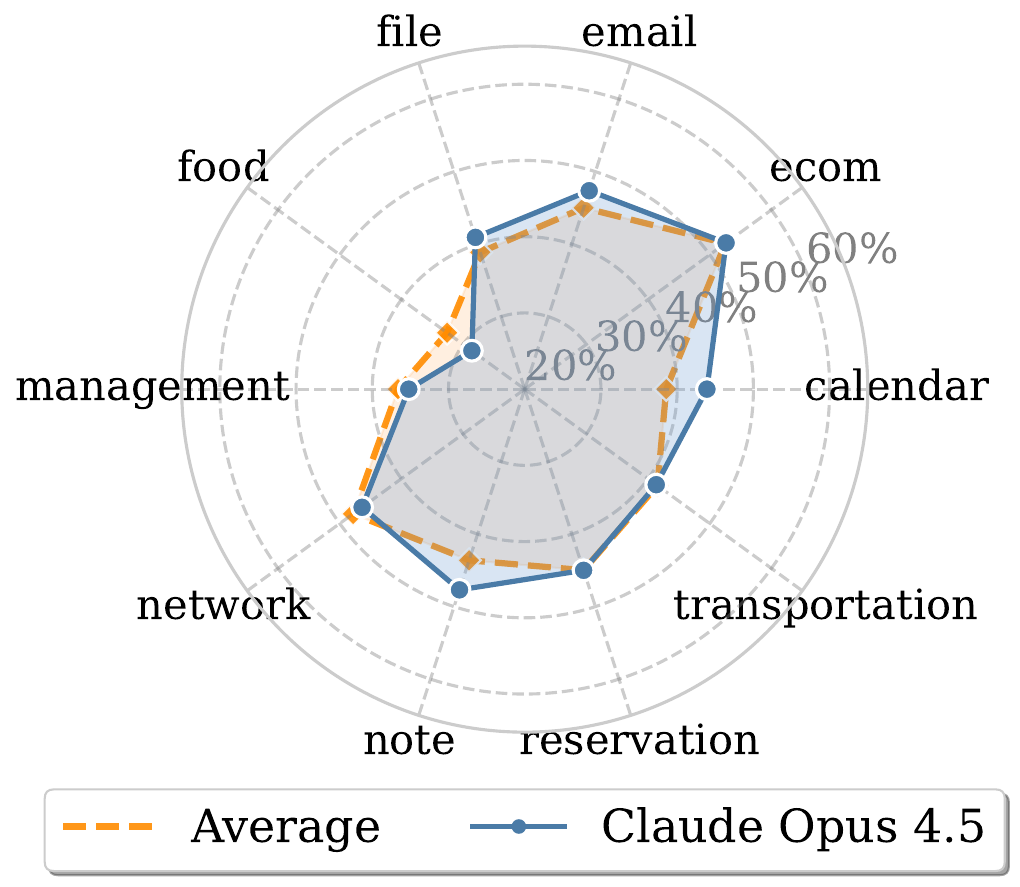}
\includegraphics[width=0.22\textwidth]{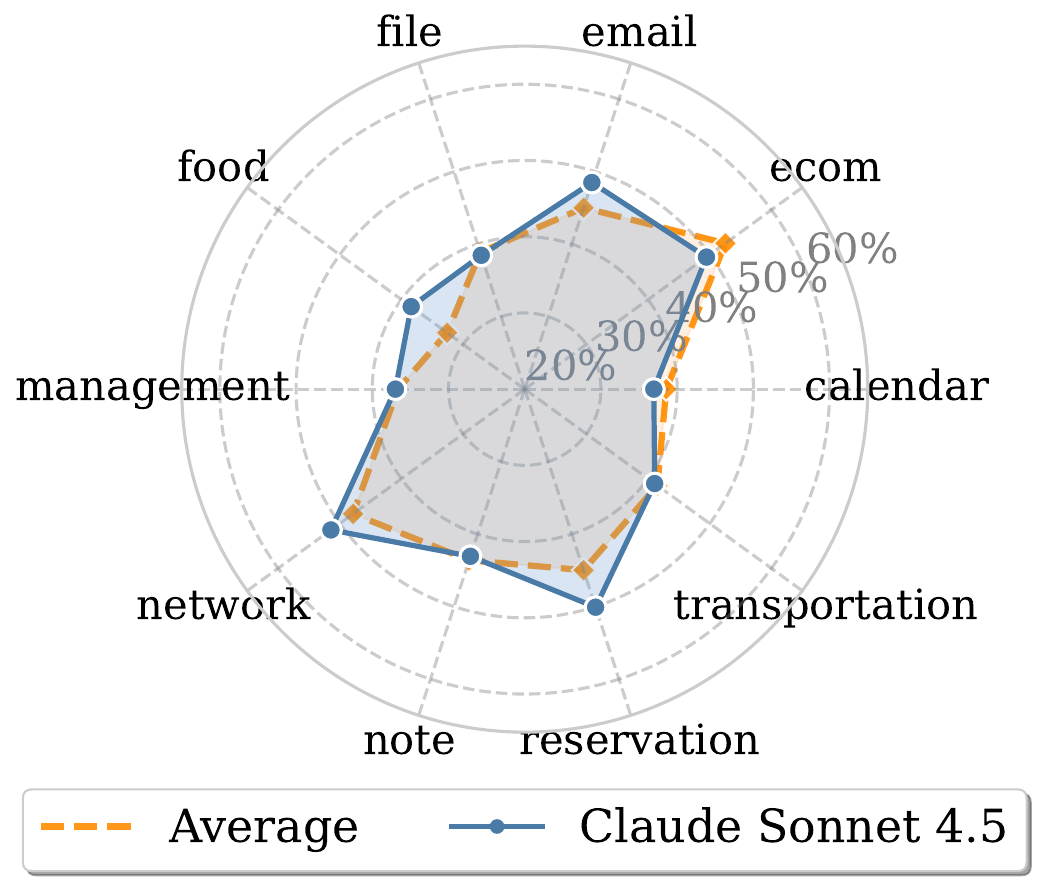}
\includegraphics[width=0.22\textwidth]{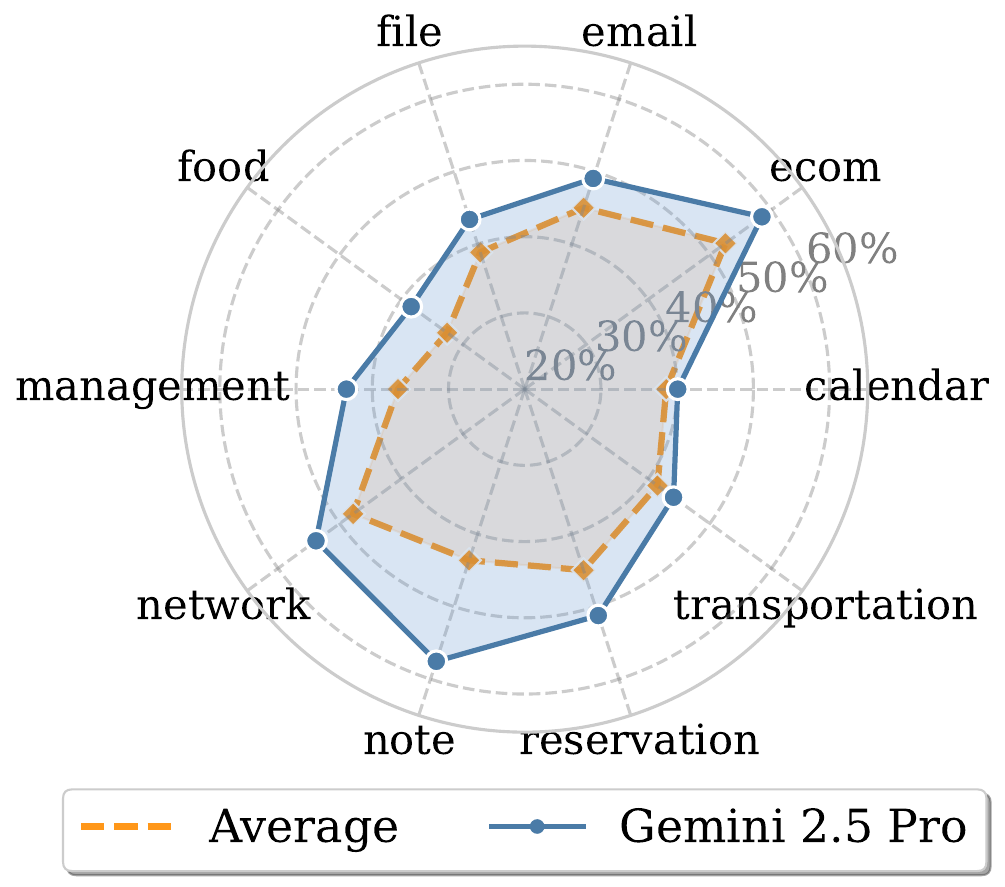}
\includegraphics[width=0.22\textwidth]{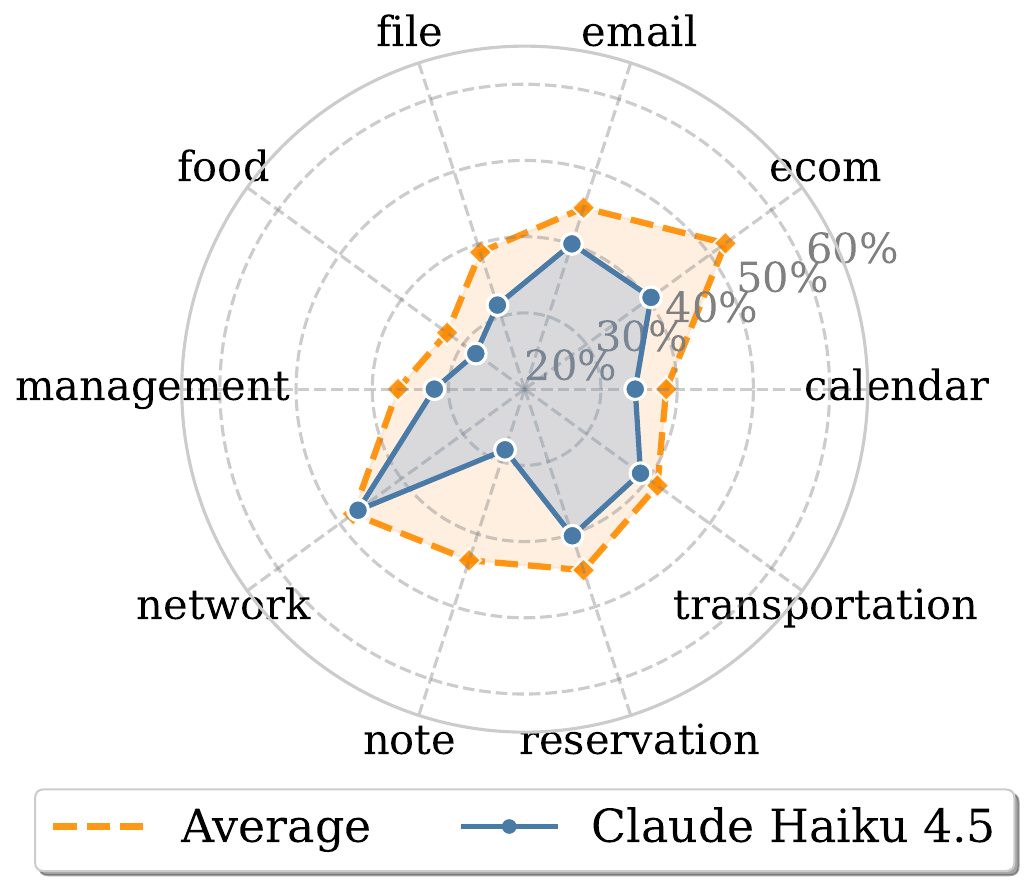}

\includegraphics[width=0.22\textwidth]{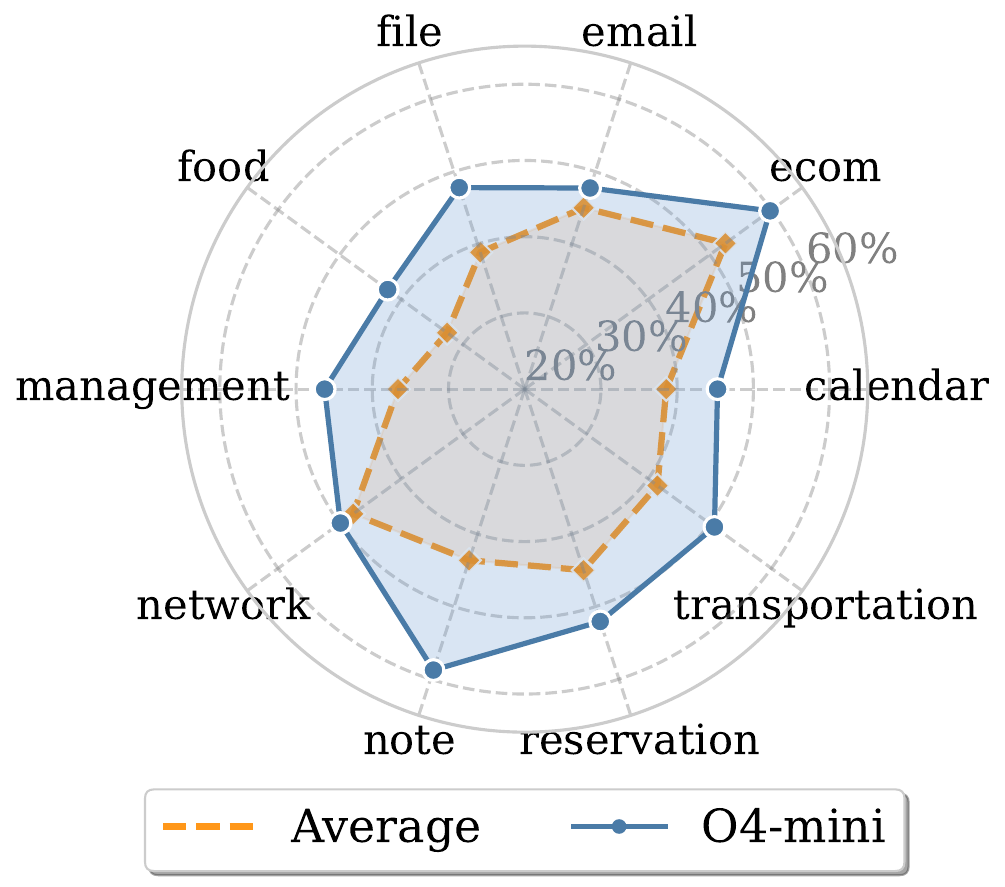}
\includegraphics[width=0.22\textwidth]{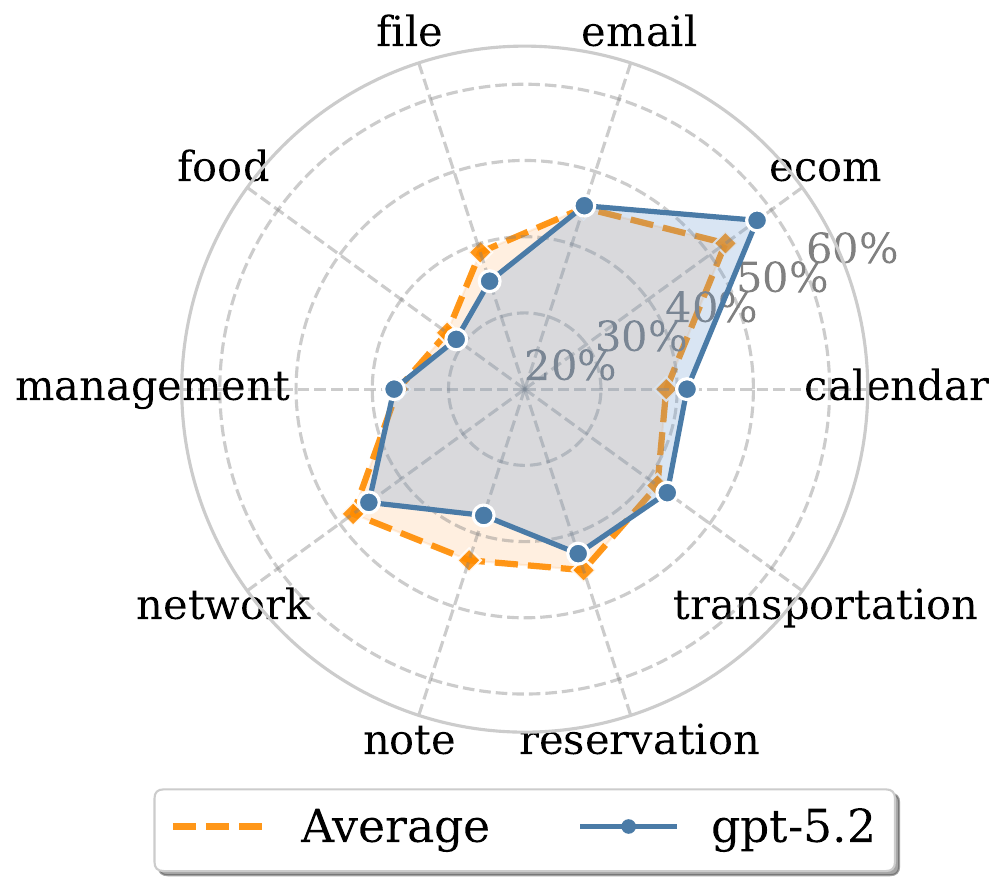}
\includegraphics[width=0.22\textwidth]{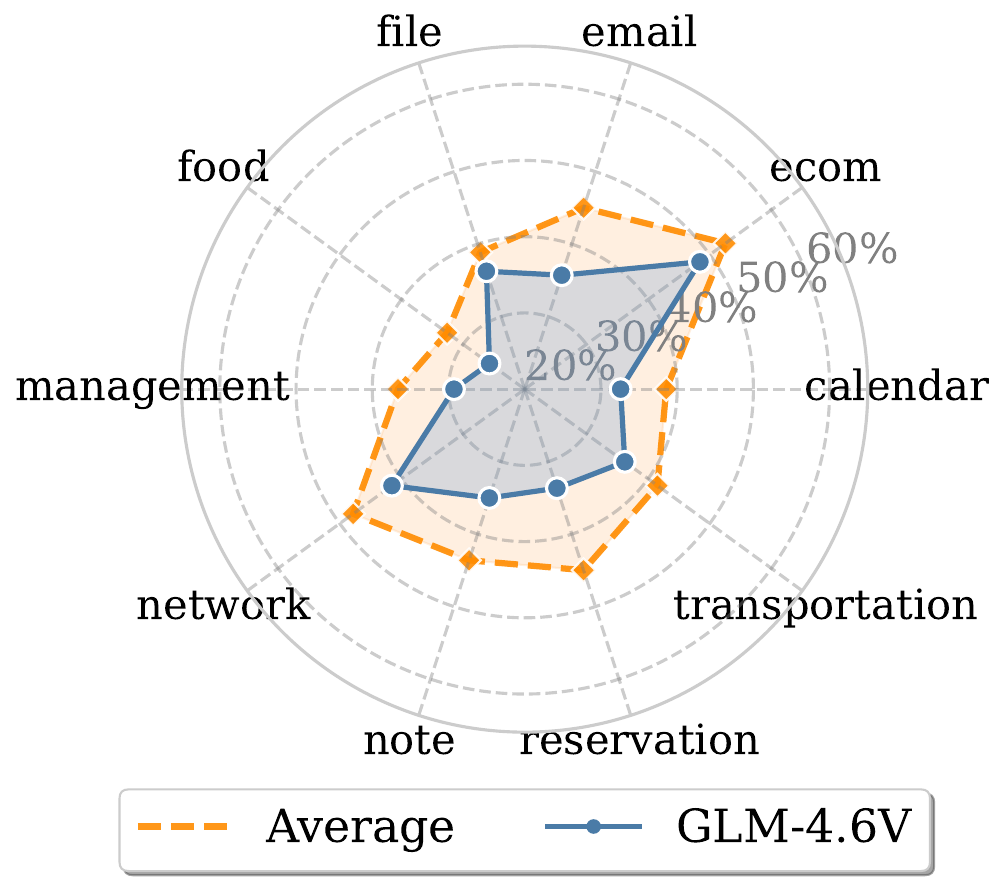}
\includegraphics[width=0.22\textwidth]{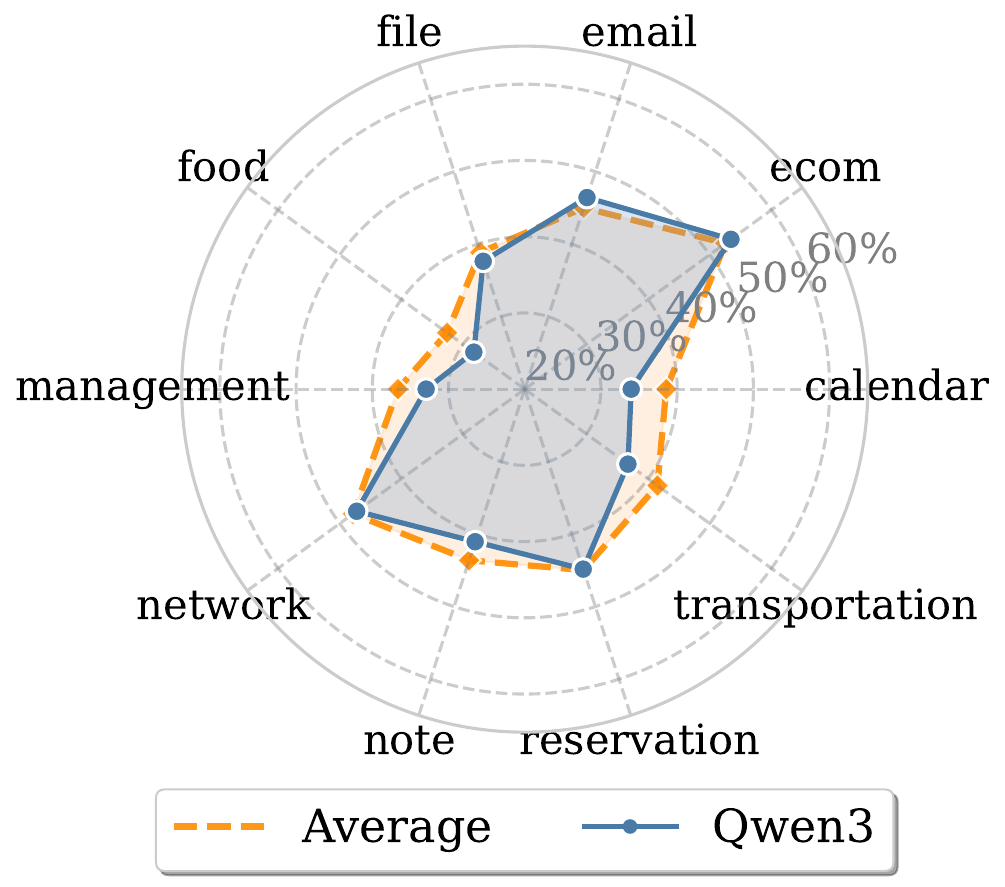}

\caption{
Radar plots showing model performance across different websites.
Each radar plot contains two curves: the performance of the specific model
and the average performance across all models.
}
\label{fig:site_radar_overview}
\end{figure*}

\begin{figure*}[t]
\centering
\includegraphics[width=0.22\textwidth]{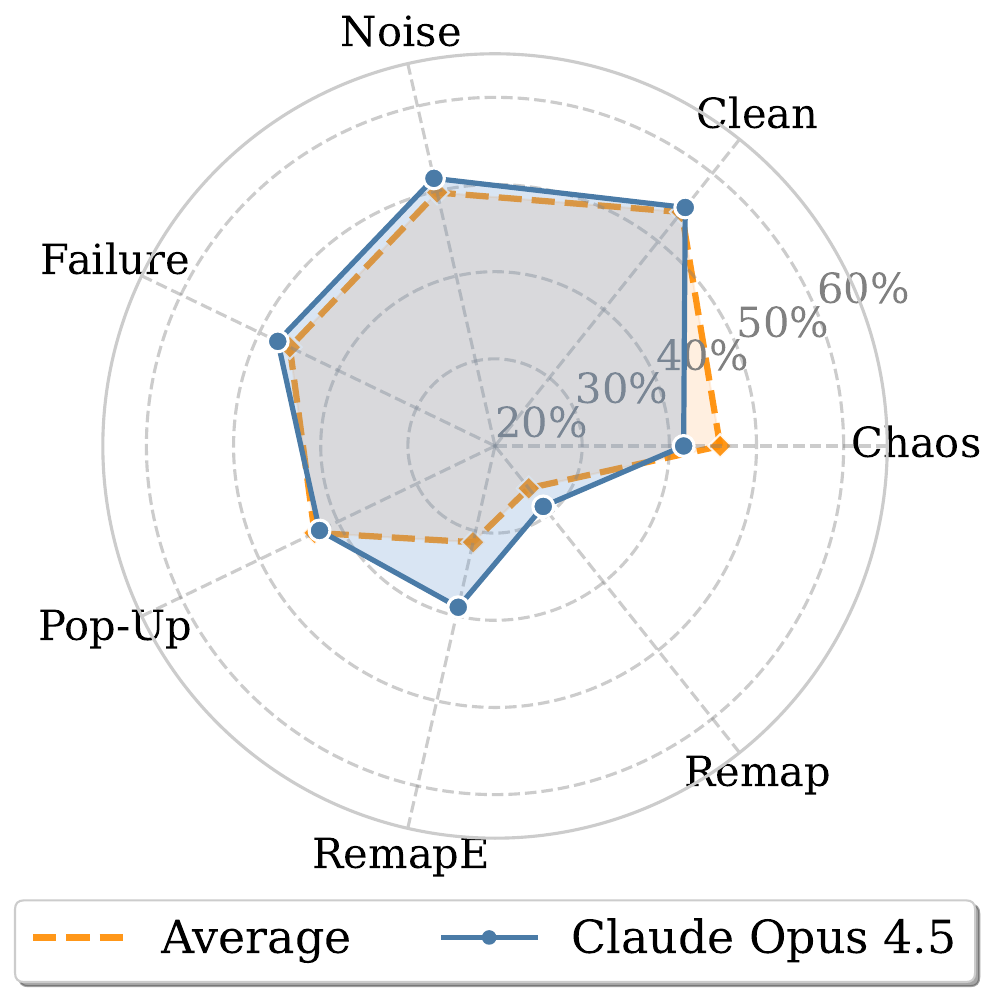}
\includegraphics[width=0.22\textwidth]{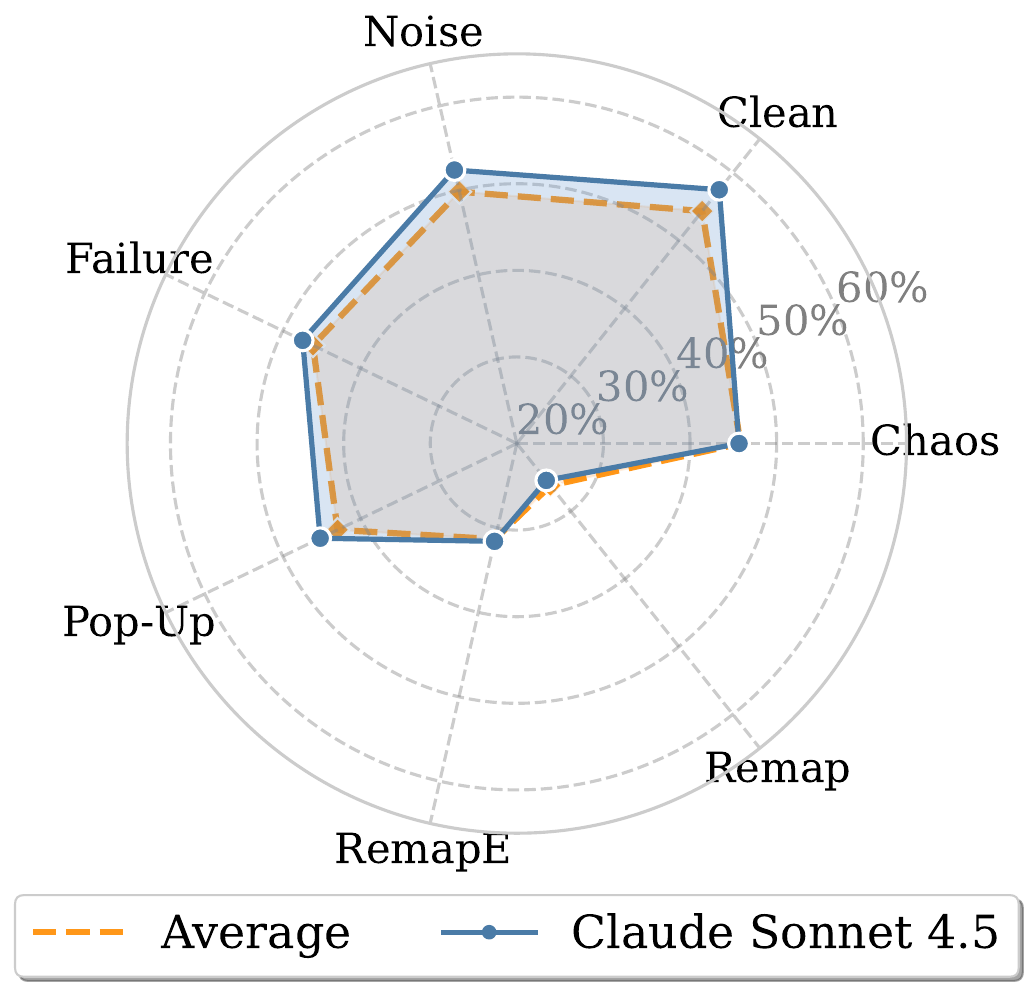}
\includegraphics[width=0.22\textwidth]{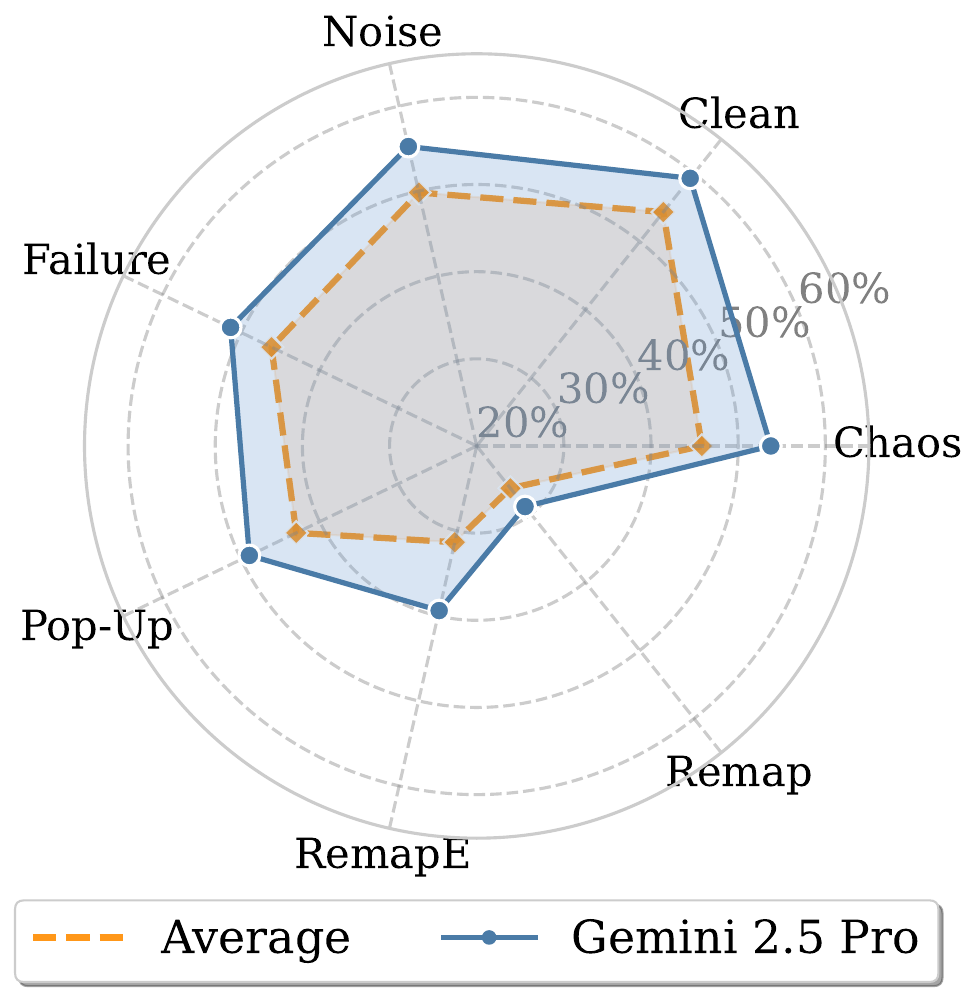}
\includegraphics[width=0.22\textwidth]{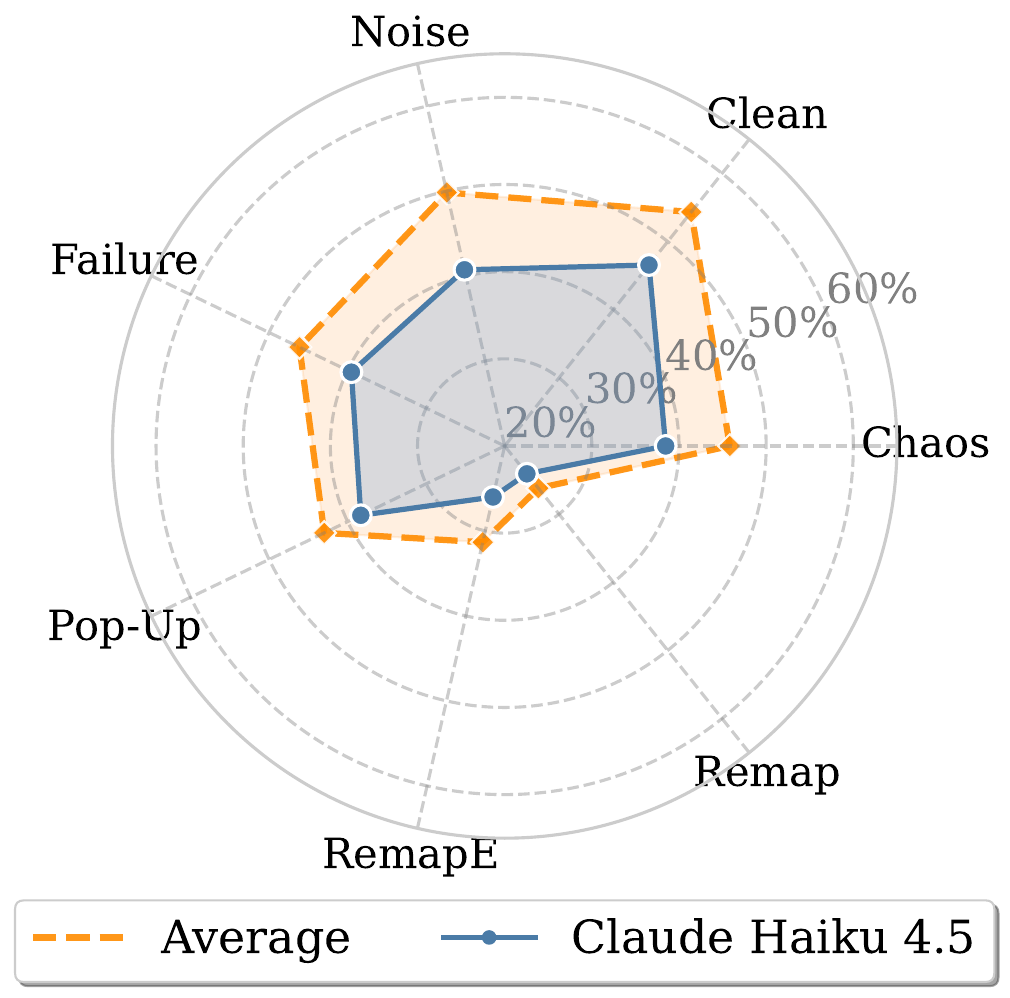}

\includegraphics[width=0.22\textwidth]{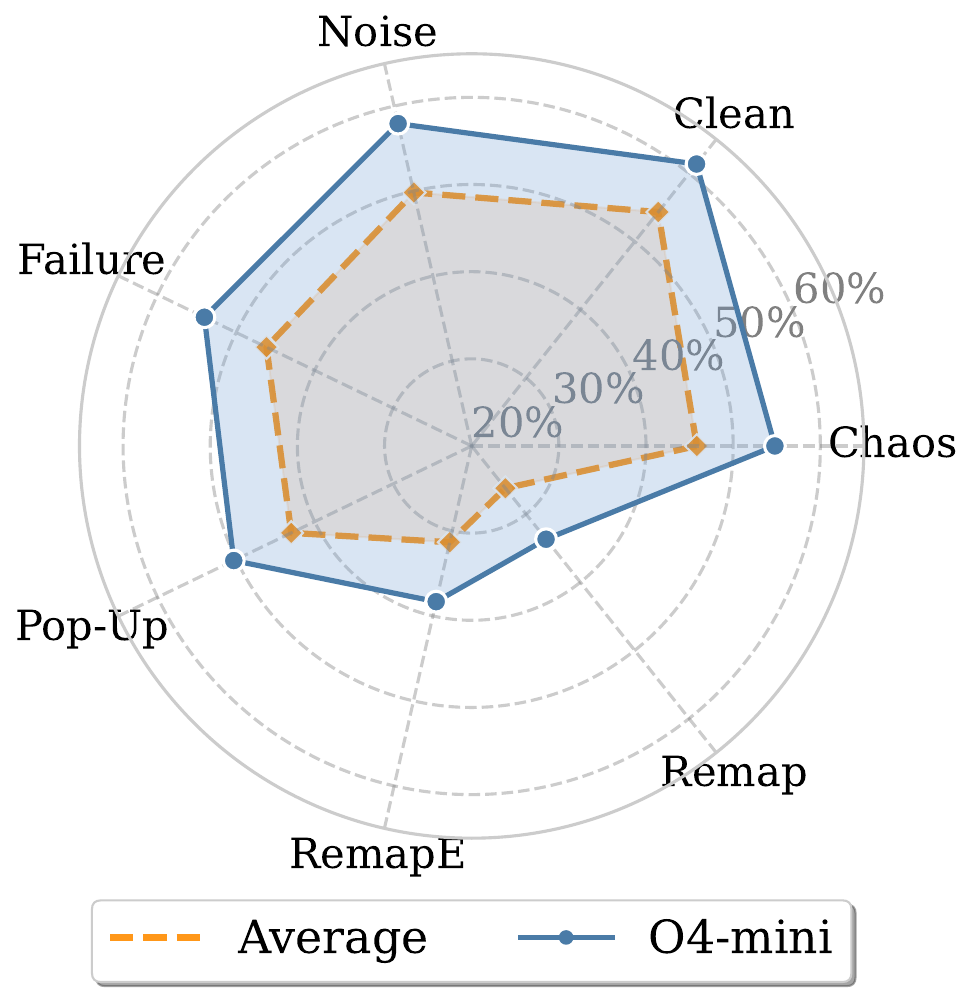}
\includegraphics[width=0.22\textwidth]{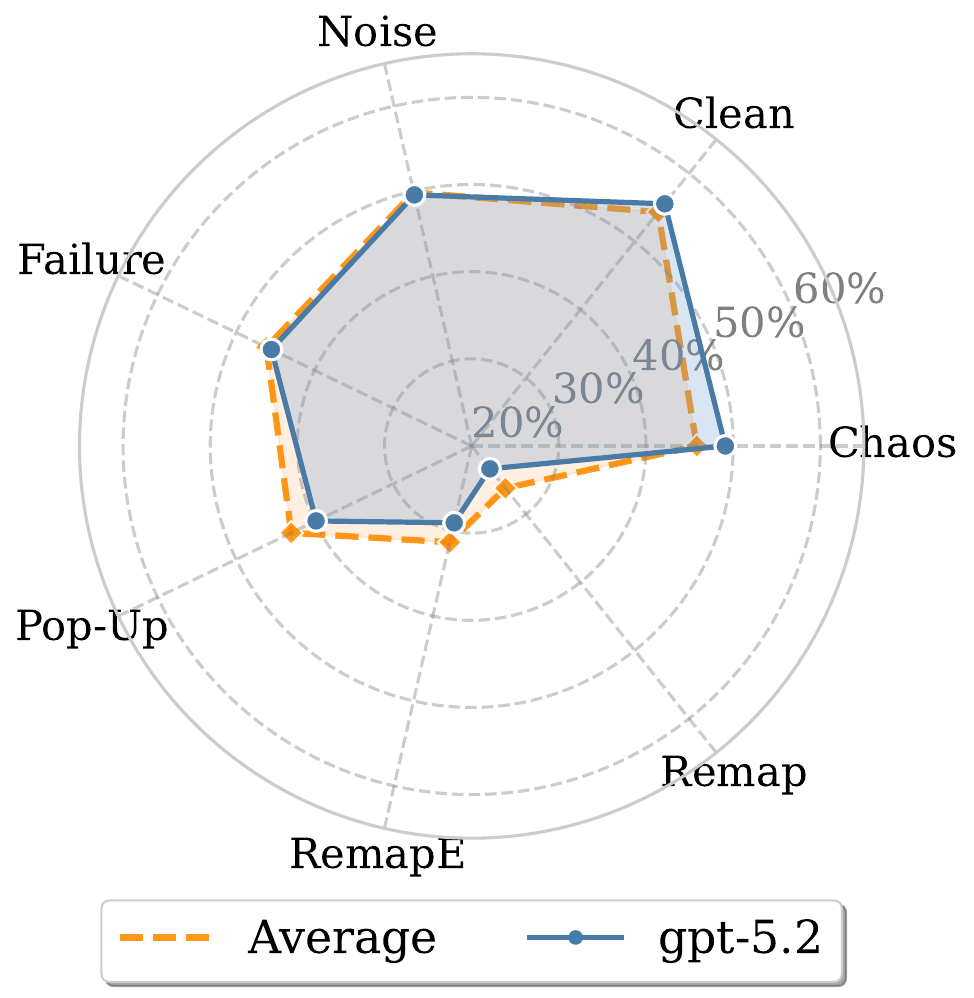}
\includegraphics[width=0.22\textwidth]{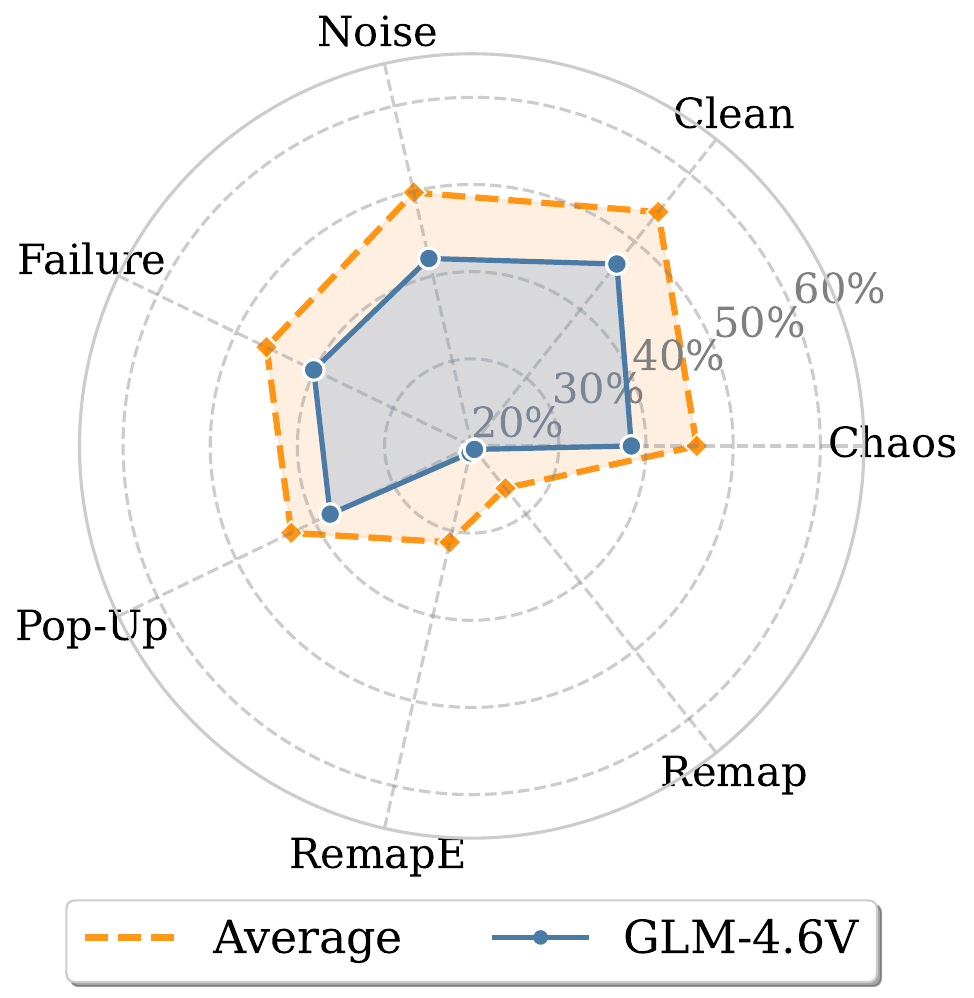}
\includegraphics[width=0.22\textwidth]{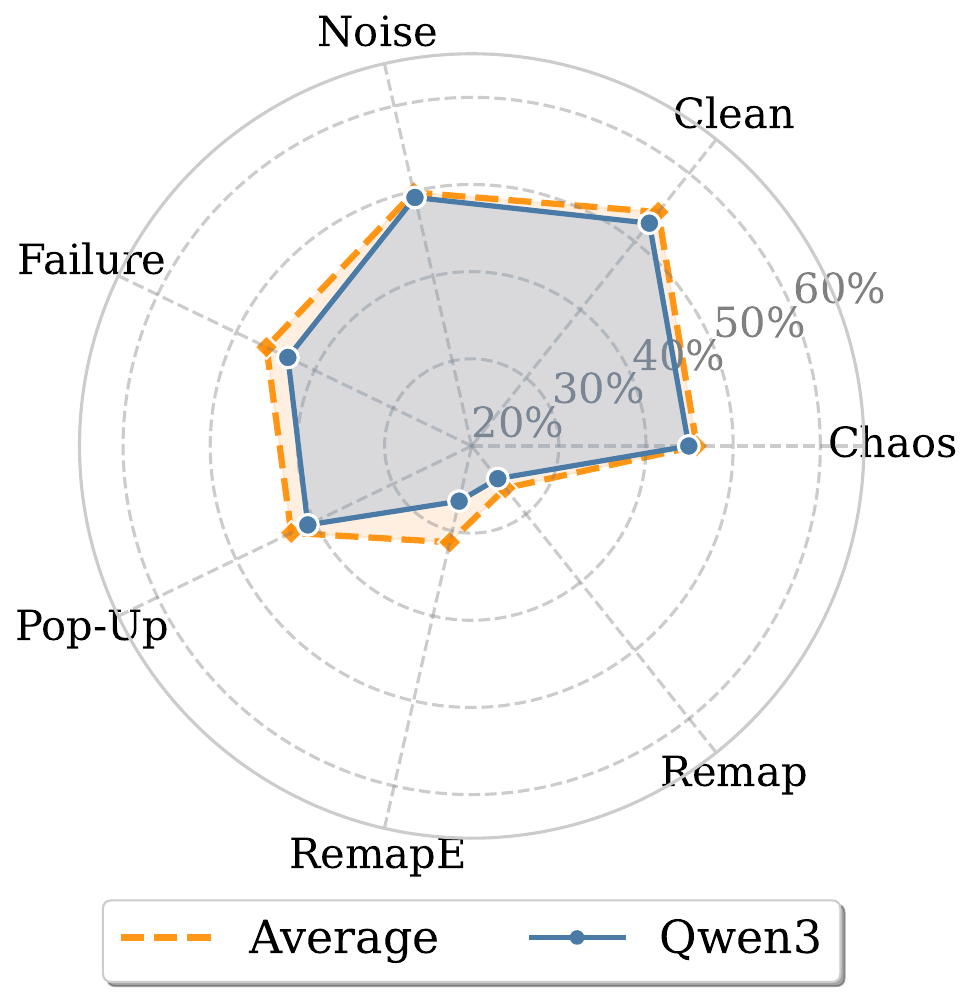}

\caption{
Radar plots showing model performance across different perturbation modes.
Each plot compares the performance of an individual model against the
average performance across all evaluated models.
}
\label{fig:mode_radar_overview}
\end{figure*}
\section{Generated Website Environments}
\label{sec:appendix_websites}

Figure~\ref{fig:Wensites} presents example screenshots of the ten generated websites used in the benchmark.
All environments are produced through the agent-driven website construction pipeline described in the main paper, which converts natural language specifications into fully functional web applications.
Unlike simplified benchmark interfaces, the generated websites maintain realistic layouts, multi-page navigation structures, and persistent internal states across interactions.
This design ensures that agent behaviors are evaluated within interaction workflows that closely resemble real-world web applications.

To cover a diverse range of realistic scenarios, the benchmark includes ten representative categories of web services spanning common online activities such as shopping, communication, scheduling, productivity, and transportation.
These environments include an e-commerce platform, food delivery service, restaurant reservation system, online notebook editor, calendar scheduling interface, email management system, list management application, professional networking platform, ride-hailing service, and cloud file management system.
Together, these domains represent widely used classes of modern web applications and provide diverse interaction patterns for evaluating web agents.

Across these environments, the websites support a broad spectrum of common web interaction primitives.
Users can navigate across multiple pages, perform keyword searches, apply filters and sorting options, interact with forms, submit multi-step requests, and update persistent data structures such as carts, reservations, schedules, messages, or file hierarchies.
Many tasks also require multi-step workflows that involve intermediate state updates before a final confirmation step.
As a result, agents must reason over both interface structure and evolving environment state in order to complete tasks successfully.

\section{Detailed Model Performance Analysis}
\label{sec:appendix_detailed_results}

This section provides detailed breakdowns of model performance across
different websites and perturbation modes.
To facilitate comparison, we report per-model radar plots showing
checkpoint pass rates under all evaluation conditions.
Each radar plot contains two curves:
(1) the performance of the specific model and
(2) the average performance across all evaluated models.
The average curve serves as a reference baseline,
allowing deviations in model-specific strengths and weaknesses
to be visually identified.
We provide two complementary views of the results.
First, we visualize performance across different websites,
highlighting how models behave under different task domains.
Second, we present performance across perturbation modes,
revealing how each model responds to different categories
of environmental disruptions.

\subsection{Performance Across Websites}
\label{sec:appendix_site_radar}

Figure~\ref{fig:site_radar_overview} presents radar plots showing
model performance across different websites.
Each plot corresponds to one evaluated model and reports the
checkpoint pass rate on all task websites.
The model-specific curve is shown together with the average
performance across all models, providing a direct reference
for identifying relative strengths and weaknesses.

\subsection{Performance Across Perturbation Modes}
\label{sec:appendix_mode_radar}

Figure~\ref{fig:mode_radar_overview} visualizes model performance
across different perturbation modes.

Each radar plot shows checkpoint pass rates under all evaluation modes,
including the clean environment and various perturbation settings.
By comparing the model-specific curve with the average baseline,
it becomes easier to identify models that exhibit stronger robustness
or higher sensitivity under particular perturbation types.

\begin{figure*}[t]
\centering
\includegraphics[width=0.23\textwidth]{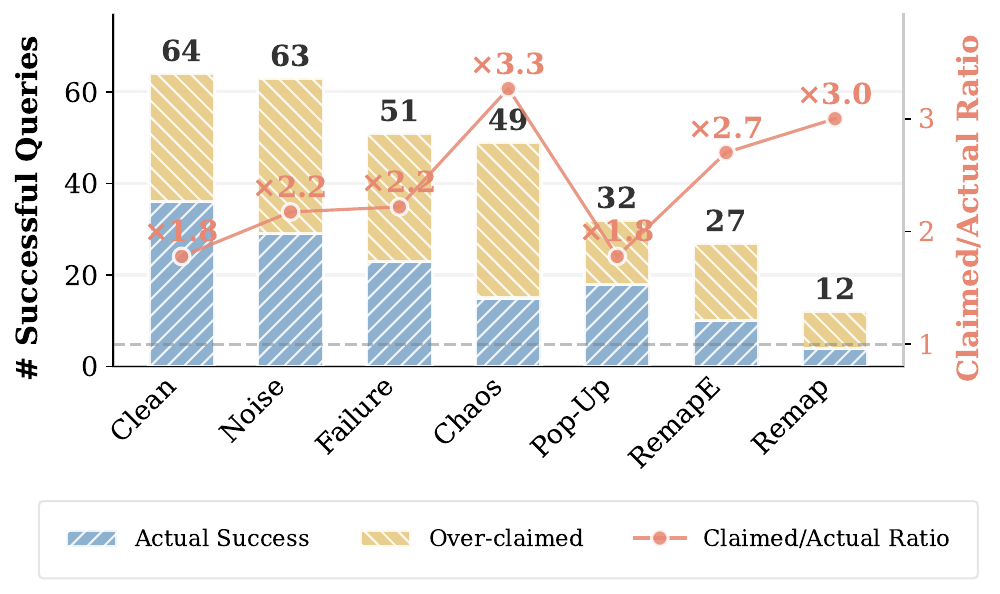}
\includegraphics[width=0.23\textwidth]{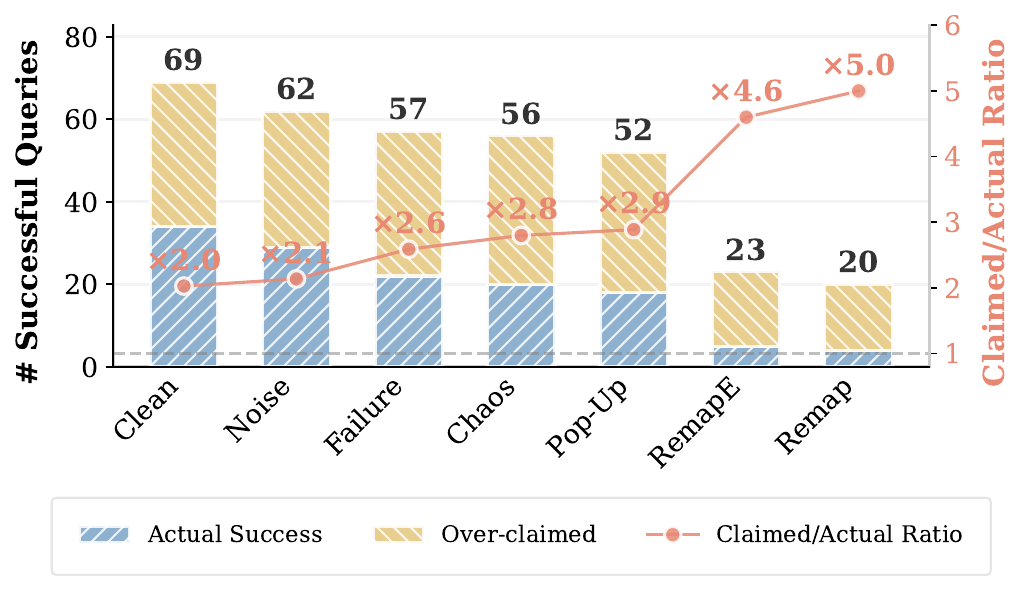}
\includegraphics[width=0.23\textwidth]{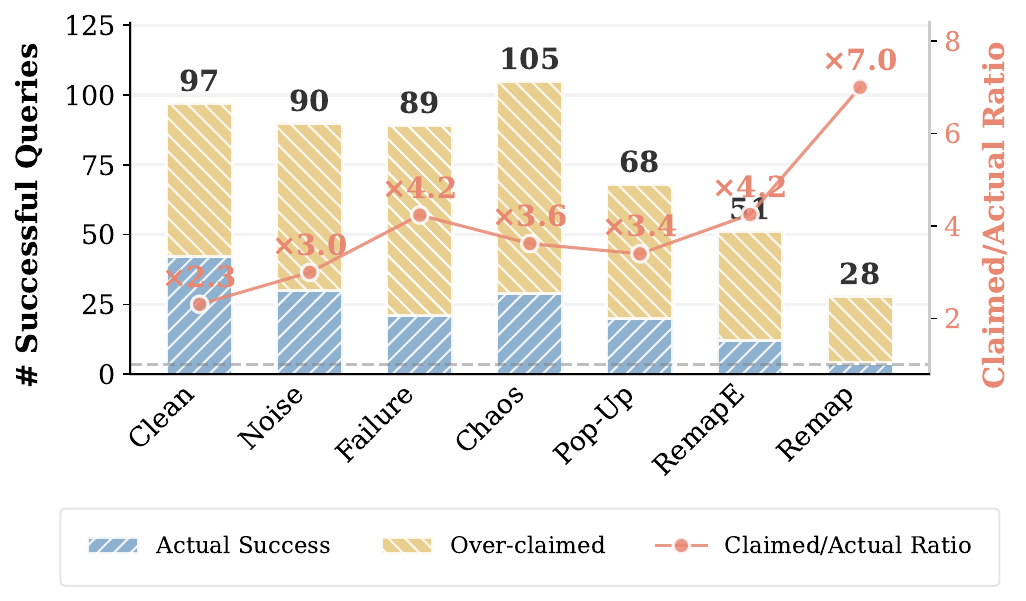}
\includegraphics[width=0.23\textwidth]{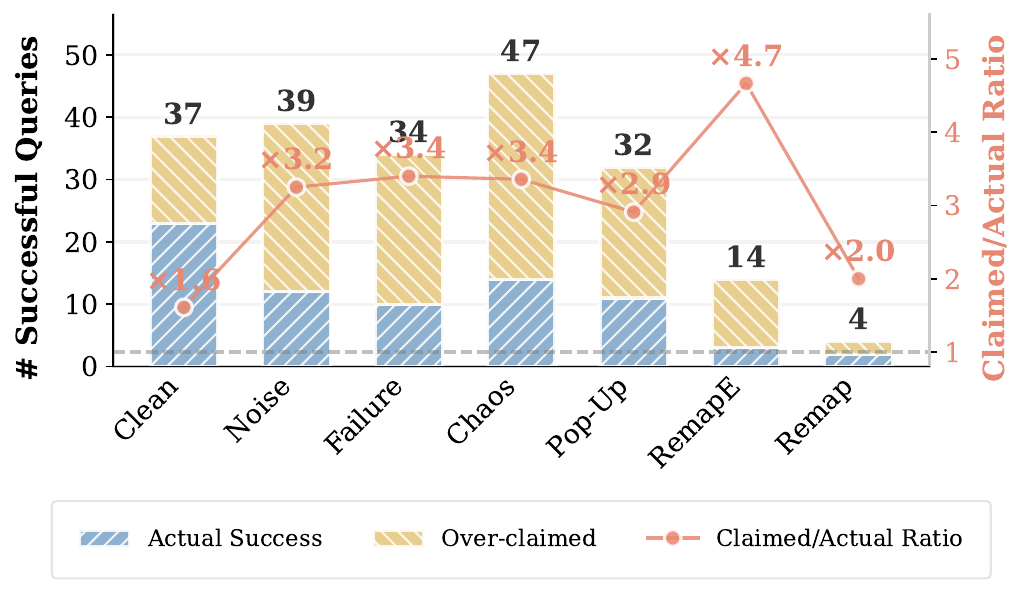}

\includegraphics[width=0.23\textwidth]{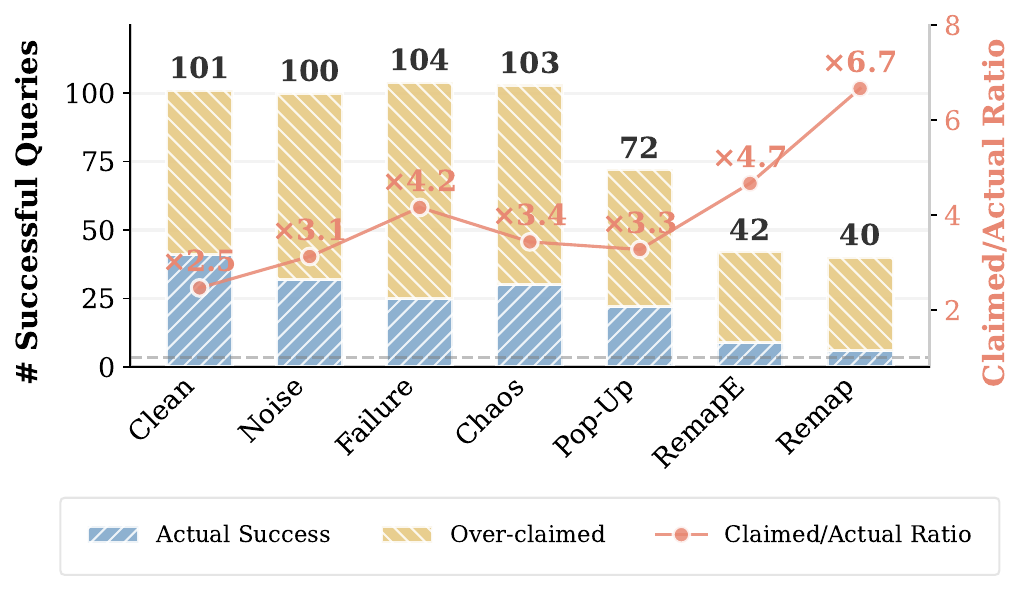}
\includegraphics[width=0.23\textwidth]{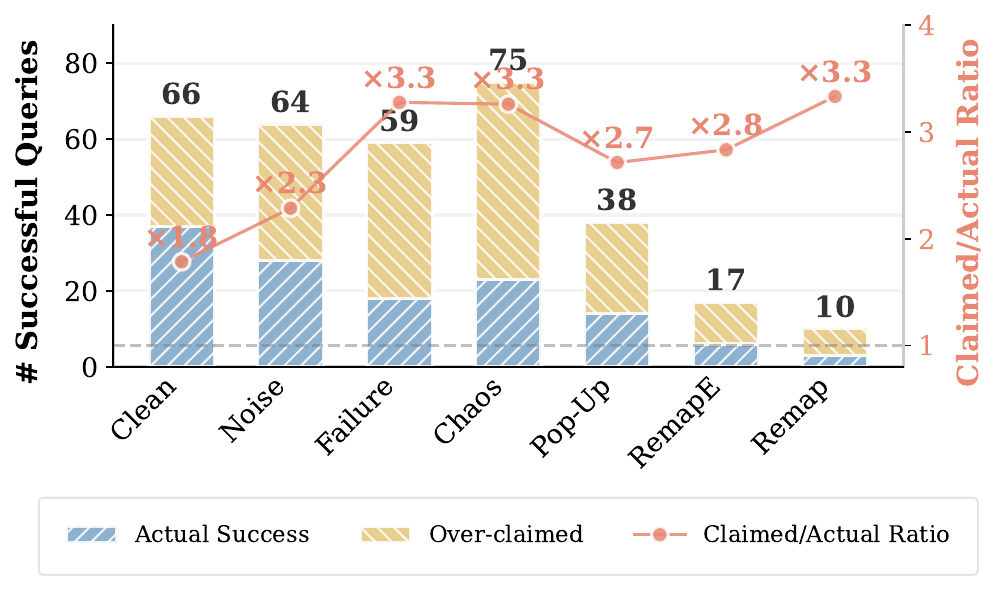}
\includegraphics[width=0.23\textwidth]{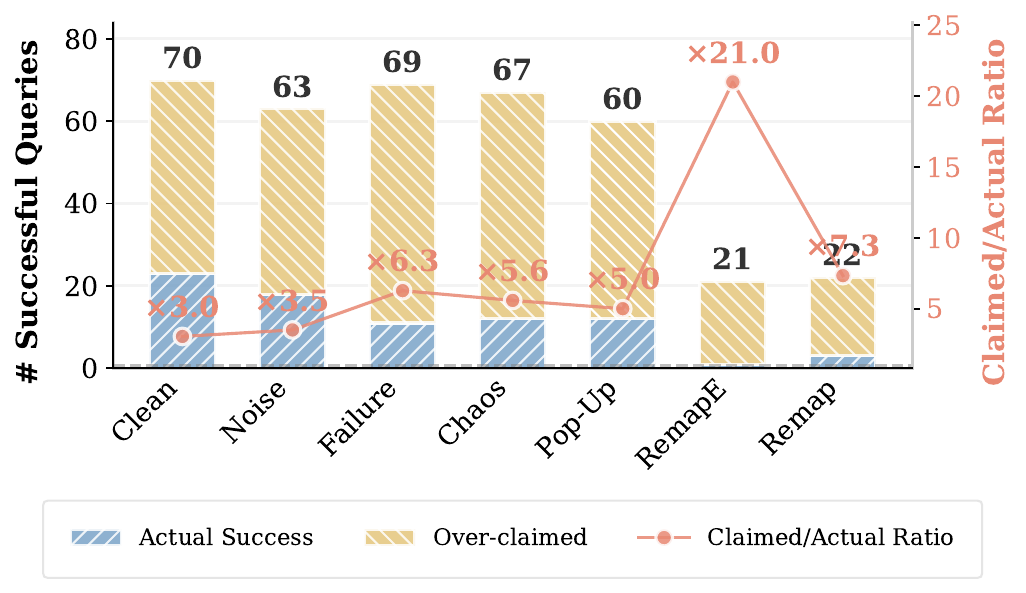}
\includegraphics[width=0.23\textwidth]{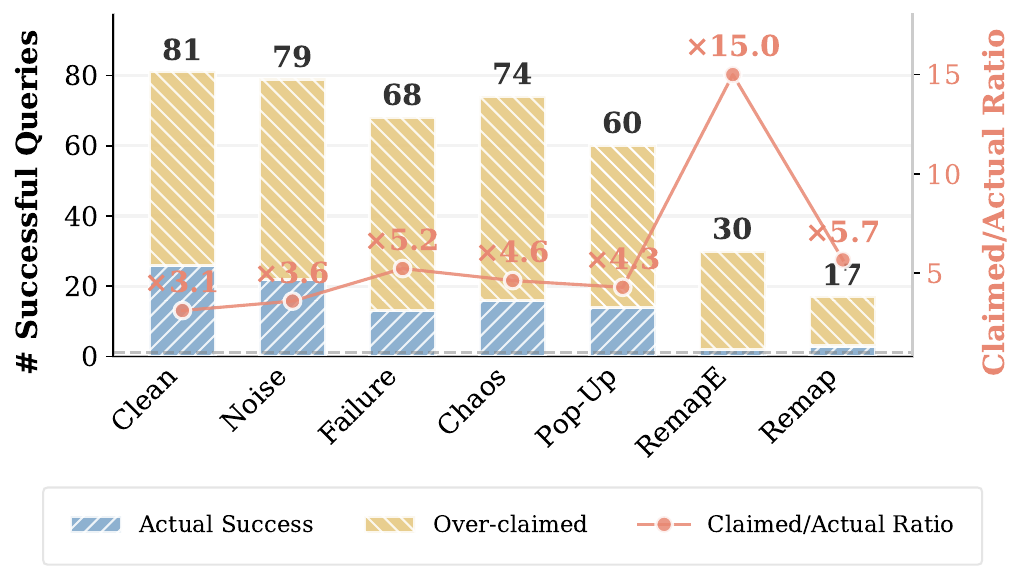}

\caption{
Model-level self-assessment reliability across different
evaluation conditions.
Each plot compares the number of queries claimed as
successful by the agent with the number of queries that
are actually successful.
The gap between the two quantities reflects the degree
of self-assessment error.
}
\label{fig:self_assessment_models}
\end{figure*}

\section{Model-Level Self-Assessment Reliability}
\label{sec:appendix_self_assessment}

To complement the aggregate analysis presented in
Section~\ref{sec:self_assessment}, we further examine
self-assessment reliability at the model level.

Figure~\ref{fig:self_assessment_models} reports the relationship
between claimed task success and actual task success for each
evaluated model under all evaluation conditions.
Each plot corresponds to one model and shows how the number
of queries claimed as successful compares with the number
of queries that are actually successful.

This model-level view reveals that the discrepancy between
claimed and actual success is not limited to a small subset
of models but appears consistently across most evaluated
systems.
While the magnitude of the gap varies across models,
environmental perturbations generally lead to systematic
overestimation of task success.
In particular, semantic perturbations often produce the
largest divergence, indicating that agents frequently fail
to recognize execution errors when interaction semantics
are altered.

\begin{figure*}[t]
\centering

\begin{subfigure}[t]{0.32\textwidth}
\centering
\includegraphics[width=\linewidth]{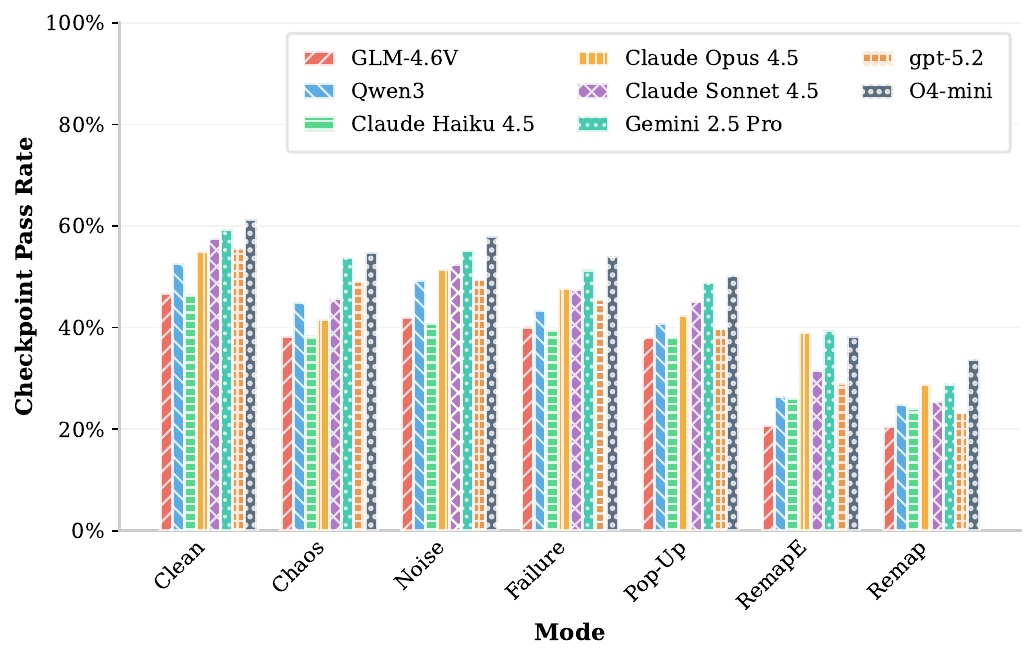}
\caption{Performance across perturbation modes.}
\label{fig:mode_group_bar}
\end{subfigure}
\hfill
\begin{subfigure}[t]{0.32\textwidth}
\centering
\includegraphics[width=\linewidth]{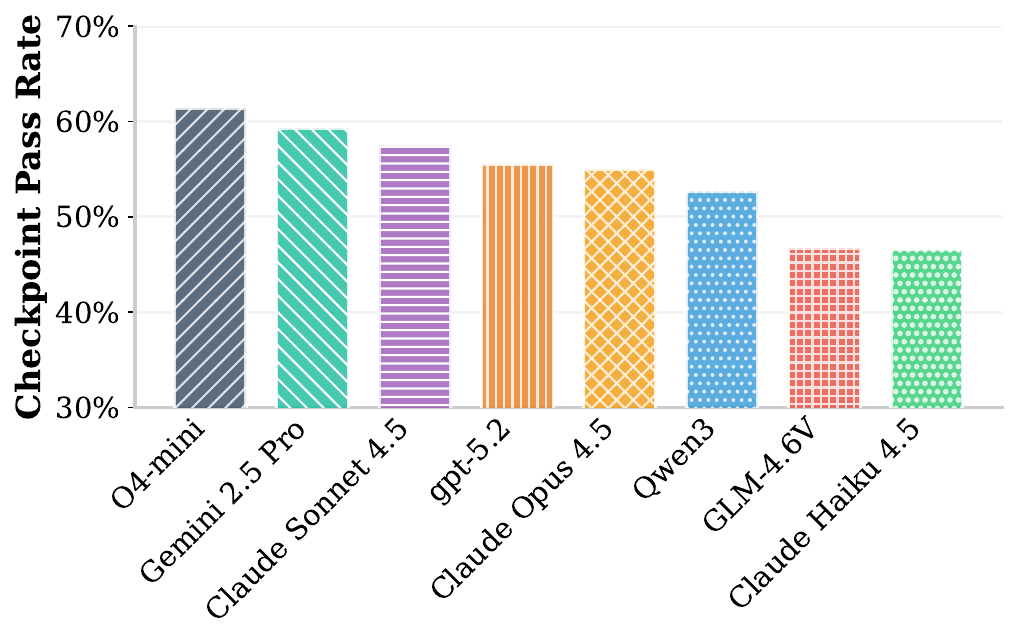}
\caption{Model performance under the clean environment.}
\label{fig:clean_bar}
\end{subfigure}
\hfill
\begin{subfigure}[t]{0.32\textwidth}
\centering
\includegraphics[width=\linewidth]{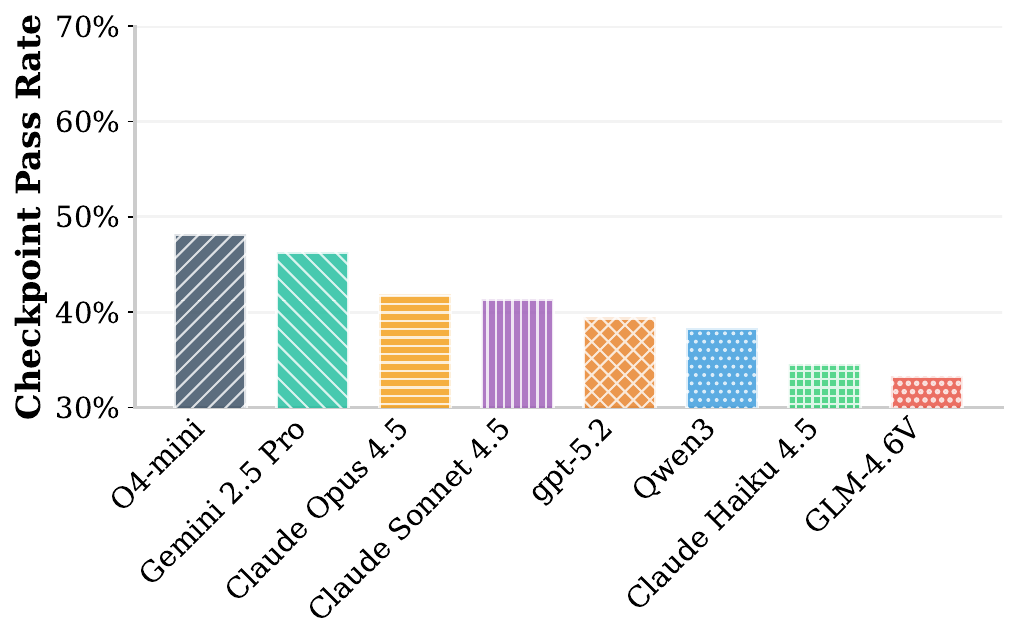}
\caption{Average performance under perturbed environments.}
\label{fig:perturbed_bar}
\end{subfigure}

\caption{
Model performance under clean and perturbed environments.
(a) Checkpoint pass rates across different perturbation modes.
(b) Model ranking under the clean environment.
(c) Average model performance across all perturbed environments.
Colors are consistent across (b) and (c), allowing ranking differences
between clean and perturbed conditions to be visually identified.
}
\label{fig:model_ranking_comparison}

\end{figure*}

\section{Model Ranking Under Clean and Perturbed Conditions}
\label{sec:appendix_model_ranking}

To further understand how environmental perturbations affect
model performance, we analyze model rankings under clean and
perturbed conditions.

Figure~\ref{fig:mode_group_bar} first reports checkpoint pass
rates across different perturbation modes for all evaluated
models.
Each group corresponds to one evaluation mode, while bars
represent different models.

We then compare model rankings between the clean environment
and perturbed environments.
Figure~\ref{fig:clean_bar} shows checkpoint pass rates under
the clean condition.
Figure~\ref{fig:perturbed_bar} reports the average checkpoint
pass rate across all perturbation modes.

To facilitate comparison, the same color is used for each
model across both figures.
This allows ranking differences between clean and perturbed
settings to be visually identified.

These comparisons reveal that model rankings under the clean
environment do not always translate to robustness under
perturbations.
Some models that perform strongly in the clean setting exhibit
larger performance degradation when environmental conditions
change, while others maintain more stable behavior across
perturbation modes.

This observation highlights the importance of evaluating web
agents beyond clean environments, as robustness to environmental
variation represents a distinct capability dimension that is
not fully captured by standard clean benchmarks.

\section{Agent Prompt Template}
\label{sec:appendix_prompt}

The following prompt template is used to instruct the agent to
interact with web environments using Playwright.
The prompt specifies the available actions, output format,
and interaction constraints.
Depending on the configuration, the agent may operate either
with DOM information or purely from screenshots.

For brevity, the list of supported actions returned by
\texttt{get\_action\_description()} is abbreviated here.

\begin{tcolorbox}[
colback=back_color,
colframe=hard_color,
title=Web Agent Prompt Template,
fonttitle=\bfseries,
breakable,
left=4pt,
right=4pt
]
\scriptsize
\begin{verbatim}
You are a web GUI automation agent that interacts with webpages
using Playwright.

Available actions include operations such as:
CLICK, TYPE, FILL, HOTKEY, WAIT, DONE, and FAIL.

Mode: <mode_name>

--------------------------------------------------

CLICK actions

When DOM information is available, prefer CSS selectors.

Selector strategy (recommended order):

1. Text-based selectors
   button:has-text("Text")
   a:has-text("Link")
   text="Exact Text"

2. Element + class
   button.primary
   div.card
   input.search-input

3. ID selector
   #element-id

4. Attribute selector
   [placeholder="Search"]
   [type="submit"]

If selector-based interaction repeatedly fails,
visual coordinates may be used as a fallback.

Selectors should remain simple and avoid deep nesting.

--------------------------------------------------

Output format

All responses must be valid JSON objects.

Example:

{
  "action_type": "CLICK",
  "parameters": {
    "selector": "button:has-text('Submit')"
  },
  "reasoning": "Click the submit button"
}

Coordinate fallback example:

{
  "action_type": "CLICK",
  "parameters": {
    "x": 200,
    "y": 250
  },
  "reasoning": "Click based on visual location"
}

--------------------------------------------------

Other actions

FILL

{
  "action_type": "FILL",
  "parameters": {
    "selector": "input[placeholder='Title']",
    "text": "example text"
  }
}

TYPE

{
  "action_type": "TYPE",
  "parameters": {
    "text": "hello"
  }
}

HOTKEY

{
  "action_type": "HOTKEY",
  "parameters": {
    "keys": "Ctrl+A"
  }
}

--------------------------------------------------

Important rules

- Always return a valid JSON object.
- Never output plain text explanations outside JSON.
- Coordinates are relative to the viewport.
- Click the center of elements whenever possible.
- If an action fails, try an alternative selector or 
adjust coordinates.
- Use DONE when the task is completed successfully.
- Use FAIL only when the task cannot be completed.
\end{verbatim}
\end{tcolorbox}

\section{Discussion}
\label{sec:discussion}

\subsection{Robustness as a Missing Dimension of Agent Capability}

Our results reveal a consistent gap between task capability and robustness in current web agents.
While modern multimodal models can achieve strong performance under clean and stable environments, their performance degrades substantially under even moderate perturbations.
This suggests that high benchmark performance does not necessarily imply reliable behavior in realistic, dynamic environments.
Instead, robustness should be considered a first-class evaluation dimension alongside task success.

\subsection{Failure to Adapt under Semantic Shifts}

A key finding of our analysis is that semantic perturbations induce disproportionately large performance degradation.
When interaction rules deviate from learned conventions, agents fail to adjust their behavior, even after repeated interaction attempts.
This indicates that current agents rely heavily on fixed interaction priors rather than dynamically grounding their actions in environment feedback.
The lack of online adaptation leads to both incorrect decisions and inefficient exploration, as agents repeatedly execute invalid actions without revising their strategies.

\subsection{Failure Patterns: Early Collapse vs. Inefficient Exploration}

By analyzing interaction steps, we identify two distinct failure patterns.
In some cases, agents terminate prematurely with fewer steps, suggesting early-stage failures caused by incorrect action assumptions.
In other cases, agents exhibit significantly longer interaction trajectories, indicating inefficient exploration without effective error correction.
These patterns highlight that failures are not solely due to increased task difficulty, but arise from systematic limitations in decision-making and adaptation mechanisms.

\subsection{Miscalibration and Overconfidence under Uncertainty}

Beyond execution failures, we observe that agents exhibit substantial miscalibration between claimed and actual success, which becomes more severe under environmental perturbations.
In particular, semantic perturbations lead to extreme overconfidence, where agents report successful task completion despite failing to achieve the intended outcome.
This suggests that current agents lack reliable internal signals for failure detection, and are unable to accurately assess their own performance under uncertainty.

\subsection{Implications for Real-World Deployment}

These findings have important implications for deploying web agents in real-world settings.
First, robustness to environmental variability is critical, as real-world interfaces are inherently dynamic and often deviate from training distributions.
Second, the lack of adaptation and self-assessment reliability may lead to silent failures, where agents terminate tasks prematurely while confidently reporting success.
Such behavior can undermine user trust and lead to downstream system errors.
Therefore, improving robustness, online adaptation, and self-assessment calibration should be key directions for future research in web agents and multimodal systems.

\begin{table}[t]
\centering
\small
\caption{Repetitive action behavior under different perturbation settings. 
Exact Repeat (\%) denotes the percentage of trajectories containing consecutive identical actions (same action type and parameters). 
Semantic perturbations lead to significantly higher repetition frequency and longer repeated sequences.}
\renewcommand{\arraystretch}{1.3}
\resizebox{\columnwidth}{!}{

\begin{tabular}{lccc}
\toprule
\textbf{Setting} & \textbf{Exact Repeat (\%)} & \textbf{Total Repeats} & \textbf{Max Repeat} \\
\midrule
Clean      & 58.4 & 1,599 & 54 \\
Chaos      & 72.5 & 1,659 & 44 \\
Failure    & 83.2 & 2,462 & 28 \\
Perturbed  & 69.1 & 1,442 & 32 \\
DOM        & 61.7 & 1,920 & 64 \\
\midrule
RemapE     & 96.6 & 4,588 & 65 \\
Remap      & 96.0 & 4,874 & 92 \\
\bottomrule
\end{tabular}
} \label{tab:repetition_analysis} \end{table}

\section{Repetitive Action Failure under Semantic Perturbations.}\label{sec:Repetitive_Action_Failure}

We further analyze repetitive action patterns to understand agent behavior under perturbations.
Table~\ref{tab:repetition_analysis} reports the proportion of trajectories containing consecutive identical actions, the total number of repeated operations, and the maximum repetition length.

Repetitive behavior is already present in the clean environment, with 58.4\% of trajectories containing exact repeated actions.
However, semantic perturbations dramatically amplify this effect.
Under RemapE and Remap, over 96\% of trajectories exhibit repeated actions, indicating that such behavior becomes nearly universal.
In addition, the total number of repeated operations increases by nearly three times compared to the clean setting, revealing a substantial escalation in repetition frequency.

More importantly, semantic perturbations induce extremely long repetition sequences.
The maximum repetition length reaches up to 92 consecutive identical actions, suggesting that agents can become trapped in persistent action loops.
Such behavior is rarely observed in the clean environment and is significantly less severe under other perturbation types.

These results indicate that semantic perturbations trigger a distinctive failure mode in which agents repeatedly execute the same action without adapting to environment feedback.
This phenomenon reflects a form of policy collapse, where agents fail to revise incorrect action hypotheses and instead remain stuck in ineffective interaction patterns.

%% file: custom.bib
@inproceedings{shi2017world,
  title={World of bits: An open-domain platform for web-based agents},
  author={Shi, Tianlin and Karpathy, Andrej and Fan, Linxi and Hernandez, Jonathan and Liang, Percy},
  booktitle={International Conference on Machine Learning},
  pages={3135--3144},
  year={2017},
  organization={PMLR}
}

@article{zhou2023webarena,
  title={Webarena: A realistic web environment for building autonomous agents},
  author={Zhou, Shuyan and Xu, Frank F and Zhu, Hao and Zhou, Xuhui and Lo, Robert and Sridhar, Abishek and Cheng, Xianyi and Ou, Tianyue and Bisk, Yonatan and Fried, Daniel and others},
  journal={arXiv preprint arXiv:2307.13854},
  year={2023}
}

@article{yan2025gui,
  title={GUI Exploration Lab: Enhancing Screen Navigation in Agents via Multi-Turn Reinforcement Learning},
  author={Yan, Haolong and Shen, Yeqing and Huang, Xin and Wang, Jia and Tan, Kaijun and Liang, Zhixuan and Li, Hongxin and Ge, Zheng and Yoshie, Osamu and Li, Si and others},
  journal={arXiv preprint arXiv:2512.02423},
  year={2025}
}

@article{pan2024webcanvas,
  title={Webcanvas: Benchmarking web agents in online environments},
  author={Pan, Yichen and Kong, Dehan and Zhou, Sida and Cui, Cheng and Leng, Yifei and Jiang, Bing and Liu, Hangyu and Shang, Yanyi and Zhou, Shuyan and Wu, Tongshuang and others},
  journal={arXiv preprint arXiv:2406.12373},
  year={2024}
}

@article{yang2025mla,
  title={Mla-trust: Benchmarking trustworthiness of multimodal llm agents in gui environments},
  author={Yang, Xiao and Chen, Jiawei and Luo, Jun and Fang, Zhengwei and Dong, Yinpeng and Su, Hang and Zhu, Jun},
  journal={arXiv preprint arXiv:2506.01616},
  year={2025}
}

@article{he2024webvoyager,
  title={Webvoyager: Building an end-to-end web agent with large multimodal models},
  author={He, Hongliang and Yao, Wenlin and Ma, Kaixin and Yu, Wenhao and Dai, Yong and Zhang, Hongming and Lan, Zhenzhong and Yu, Dong},
  journal={arXiv preprint arXiv:2401.13919},
  year={2024}
}

@article{garg2025real,
  title={Real: Benchmarking autonomous agents on deterministic simulations of real websites},
  author={Garg, Divyansh and VanWeelden, Shaun and Caples, Diego and Draguns, Andis and Ravi, Nikil and Putta, Pranav and Garg, Naman and Abraham, Tomas and Lara, Michael and Lopez, Federico and others},
  journal={arXiv preprint arXiv:2504.11543},
  year={2025}
}

@article{kara2025warex,
  title={WAREX: Web Agent Reliability Evaluation on Existing Benchmarks},
  author={Kara, Su and Faisal, Fazle and Nath, Suman},
  journal={arXiv preprint arXiv:2510.03285},
  year={2025}
}

@article{wei2025webagent,
  title={Webagent-r1: Training web agents via end-to-end multi-turn reinforcement learning},
  author={Wei, Zhepei and Yao, Wenlin and Liu, Yao and Zhang, Weizhi and Lu, Qin and Qiu, Liang and Yu, Changlong and Xu, Puyang and Zhang, Chao and Yin, Bing and others},
  journal={arXiv preprint arXiv:2505.16421},
  year={2025}
}

@inproceedings{ning2025survey,
  title={A survey of webagents: Towards next-generation ai agents for web automation with large foundation models},
  author={Ning, Liangbo and Liang, Ziran and Jiang, Zhuohang and Qu, Haohao and Ding, Yujuan and Fan, Wenqi and Wei, Xiao-yong and Lin, Shanru and Liu, Hui and Yu, Philip S and others},
  booktitle={Proceedings of the 31st ACM SIGKDD Conference on Knowledge Discovery and Data Mining V. 2},
  pages={6140--6150},
  year={2025}
}

@article{violations1,
  title={Non-flaky and nearly optimal time-based treatment of asynchronous wait web tests},
  author={Pei, Yu and Sohn, Jeongju and Habchi, Sarra and Papadakis, Mike},
  journal={ACM Transactions on Software Engineering and Methodology},
  volume={34},
  number={2},
  pages={1--29},
  year={2025},
  publisher={ACM New York, NY}
}

@article{violations2,
  title={An Empirical Study of Web Visual Flakiness: Characterisation and Fix Strategies},
  author={Pei, Yu and Sohn, Jeongju and Papadakis, Mike},
  journal={Journal of Systems and Software},
  pages={112826},
  year={2026},
  publisher={Elsevier}
}

@incollection{violations3,
  title={Cumulative Layout Shift},
  author={Edgar, Matthew},
  booktitle={Speed Metrics Guide: Choosing the Right Metrics to Use When Evaluating Websites},
  pages={157--180},
  year={2024},
  publisher={Springer}
}

@article{popup,
  title={Popsweeper: Automatically detecting and resolving app-blocking pop-ups to assist automated mobile gui testing},
  author={Guo, Linqiang and Liu, Wei and Heng, Yi Wen and Wang, Yang and others},
  journal={arXiv preprint arXiv:2412.02933},
  year={2024}
}

@article{danger,
  title={The hidden dangers of browsing ai agents},
  author={Mudryi, Mykyta and Chaklosh, Markiyan and W{\u{A}}{\l}jcik, Grzegorz},
  journal={arXiv preprint arXiv:2505.13076},
  year={2025}
}

@inproceedings{dompert,
  title={Ghosts in the Markup: Techniques to Fight Large Language Model-Powered Web Scrapers},
  author={Brach, William and Petrik, Matej and Ko{\v{s}}t’{\'a}l, Kristi{\'a}n and Ries, Michal},
  booktitle={2025 37th Conference of Open Innovations Association (FRUCT)},
  pages={37--46},
  year={2025},
  organization={IEEE}
}

@article{sarabamoun2025special,
  title={Special-Character Adversarial Attacks on Open-Source Language Model},
  author={Sarabamoun, Ephraiem},
  journal={arXiv preprint arXiv:2508.14070},
  year={2025}
}

@inproceedings{sepert1,
  title={Explicating" Implicit Interaction" An examination of the concept and challenges for research},
  author={Serim, Bar{\i}{\c{s}} and Jacucci, Giulio},
  booktitle={Proceedings of the 2019 chi conference on human factors in computing systems},
  pages={1--16},
  year={2019}
}

@article{wei2025browsecomp,
  title={Browsecomp: A simple yet challenging benchmark for browsing agents},
  author={Wei, Jason and Sun, Zhiqing and Papay, Spencer and McKinney, Scott and Han, Jeffrey and Fulford, Isa and Chung, Hyung Won and Passos, Alex Tachard and Fedus, William and Glaese, Amelia},
  journal={arXiv preprint arXiv:2504.12516},
  year={2025}
}

@article{li2025mmbrowsecomp,
  title={Mm-browsecomp: A comprehensive benchmark for multimodal browsing agents},
  author={Li, Shilong and Bu, Xingyuan and Wang, Wenjie and Liu, Jiaheng and Dong, Jun and He, Haoyang and Lu, Hao and Zhang, Haozhe and Jing, Chenchen and Li, Zhen and others},
  journal={arXiv preprint arXiv:2508.13186},
  year={2025}
}

@article{deng2023mind2web,
  title={Mind2web: Towards a generalist agent for the web},
  author={Deng, Xiang and Gu, Yu and Zheng, Boyuan and Chen, Shijie and Stevens, Sam and Wang, Boshi and Sun, Huan and Su, Yu},
  journal={Advances in Neural Information Processing Systems},
  volume={36},
  pages={28091--28114},
  year={2023}
}

@article{yao2022webshop,
  title={Webshop: Towards scalable real-world web interaction with grounded language agents},
  author={Yao, Shunyu and Chen, Howard and Yang, John and Narasimhan, Karthik},
  journal={Advances in Neural Information Processing Systems},
  volume={35},
  pages={20744--20757},
  year={2022}
}

@article{lyu2025deepshop,
  title={Deepshop: A benchmark for deep research shopping agents},
  author={Lyu, Yougang and Zhang, Xiaoyu and Yan, Lingyong and de Rijke, Maarten and Ren, Zhaochun and Chen, Xiuying},
  journal={arXiv preprint arXiv:2506.02839},
  year={2025}
}

@article{peeters2025webmall,
  title={WebMall--A Multi-Shop Benchmark for Evaluating Web Agents},
  author={Peeters, Ralph and Steiner, Aaron and Schwarz, Luca and Yuya Caspary, Julian and Bizer, Christian},
  journal={arXiv e-prints},
  pages={arXiv--2508},
  year={2025}
}

@article{anupam2025browserarena,
  title={BrowserArena: Evaluating LLM Agents on Real-World Web Navigation Tasks},
  author={Anupam, Sagnik and Brown, Davis and Li, Shuo and Wong, Eric and Hassani, Hamed and Bastani, Osbert},
  journal={arXiv preprint arXiv:2510.02418},
  year={2025}
}

@article{shao2026baldro,
  title={BalDRO: A Distributionally Robust Optimization based Framework for Large Language Model Unlearning},
  author={Shao, Pengyang and Zhai, Naixin and Chen, Lei and Yang, Yonghui and Zhu, Fengbin and Yang, Xun and Wang, Meng},
  journal={arXiv preprint arXiv:2601.09172},
  year={2026}
}

@article{liu2025debate,
  title={Debate over Mixed-knowledge: A Robust Multi-Agent Framework for Incomplete Knowledge Graph Question Answering},
  author={Liu, Jilong and Shao, Pengyang and Qin, Wei and Liu, Fei and Yang, Yonghui and Hong, Richang},
  journal={arXiv e-prints},
  pages={arXiv--2511},
  year={2025}
}

@article{zhai2026maximizing,
  title={Maximizing Local Entropy Where It Matters: Prefix-Aware Localized LLM Unlearning},
  author={Zhai, Naixin and Shao, Pengyang and Zheng, Binbin and Shen, Fei and Bai, Long and Yang, Xun},
  journal={arXiv preprint arXiv:2601.03190},
  year={2026}
}
